\documentclass[]{elsart}

\usepackage{icarus}

\usepackage{natbib}
\usepackage{aas_macros}
\usepackage{array}

\usepackage{hyperref}
\usepackage{csquotes}
\hypersetup{colorlinks=true,linkcolor=blue,citecolor=blue,urlcolor=blue}

\usepackage{adjustbox}
\usepackage{rotating}

\bibpunct{(}{)}{;}{a}{,}{,}

\usepackage{graphicx}

\begin{document}


\title{The~nitrogen~cycles~on~Pluto~over~seasonal~and astronomical~timescales}
\begin{frontmatter}

\author[a,b]{T. Bertrand}, 
\author[a]{F. Forget},
\author[b]{O.M. Umurhan},
\author[c]{W.M. Grundy},
\author[d]{B. Schmitt},
\author[e,f]{S. Protopapa},
\author[f]{A.M. Zangari},
\author[b]{O.L. White},
\author[g]{P.M. Schenk},
\author[f]{K.N. Singer},
\author[f]{A. Stern},
\author[h]{H.A. Weaver},
\author[f]{L.A. Young},
\author[b]{K. Ennico},
\author[f]{ C.B. Olkin},
\author[]{the New Horizons Science Team},

\address[a]{Laboratoire de M\'et\'orologie Dynamique, IPSL, Sorbonne Universités, UPMC Univ Paris 06, CNRS, 4 place Jussieu, 75005 Paris, France.}
\address[b]{National Aeronautics and Space Administration (NASA), Ames Research Center, Space Science Division, Moffett Field, CA 94035, United States }
\address[c]{Lowell Observatory, Flagstaff, AZ, United States}
\address[d]{Université Grenoble Alpes, CNRS, Institut de Plan\'etologie et Astrophysique de Grenoble, F-38000 Grenoble, France}
\address[e]{University of Maryland, Department of Astronomy, College Park, MD 20742, United States}
\address[f]{Southwest Research Institute, Boulder, CO 80302, United States}
\address[g]{Lunar and Planetary Institute, 3600 Bay Area Blvd. Houston, TX 77058, United States }
\address[h]{Johns Hopkins University Applied Physics Laboratory, Laurel, MD 20723, United States}

\begin{center}
\scriptsize
Copyright \copyright\ 2005, 2006 Ross A. Beyer, David P. O'Brien, Paul
Withers, and Gwen Bart
\end{center}

\end{frontmatter}

\begin{flushleft}
Number of pages: \pageref{lastpage} \\
Number of tables: \ref{lasttable}\\
Number of figures: \ref{lastfig}\\
\end{flushleft}

\begin{pagetwo}{}

Tanguy Bertrand \\
Laboratoire de M\'et\'orologie Dynamique, CNRS/UPMC (France)\\

Email: tanguy.bertrand@lmd.jussieu.fr\\

\end{pagetwo}

\begin{abstract}
Pluto's landscape is shaped by the endless condensation and sublimation cycles of the volatile ices covering its surface. In particular, the Sputnik Planitia ice sheet, which is thought to be the main reservoir of nitrogen ice, displays a large diversity of terrains, with bright and dark plains, small pits and troughs, topographic depressions and evidences of recent and past glacial flows. Outside Sputnik Planitia, New Horizons also revealed numerous nitrogen ice deposits, in the eastern side of Tombaugh Regio and at mid-northern latitudes.

These observations suggest a complex history involving volatile and glacial processes occurring on different timescales. 
We present numerical simulations of volatile transport on Pluto performed with a model designed to simulate the nitrogen cycle over millions of years, taking into account the changes of obliquity, solar longitude of perihelion and eccentricity as experienced by Pluto. Using this model, we first explore how the volatile and glacial activity of nitrogen within Sputnik Planitia has been impacted by the diurnal, seasonal and astronomical cycles of Pluto. 
Results show that the obliquity dominates the N$_2$ cycle and that over one obliquity cycle, the latitudes of Sputnik Planitia between 25$^{\circ}$S-30$^{\circ}$N are dominated by N$_2$ condensation, while the northern regions between 30$^{\circ}$N-50$^{\circ}$N are dominated by  N$_2$ sublimation. We find that a net amount of 1 km of ice has sublimed at the northern edge of Sputnik Planitia during the last 2 millions of years. It must have been compensated by a viscous flow of the thick ice sheet.
By comparing these results with the observed geology of Sputnik Planitia, we can relate the formation of the small pits and the brightness of the ice at the center of Sputnik Planitia to the sublimation and condensation of ice occurring at the annual timescale, while the glacial flows at its eastern edge and the erosion of the water ice mountains all around the ice sheet are instead related to the astronomical timescale.
We also perform simulations including a glacial flow scheme which shows that the Sputnik Planitia ice sheet is currently at its minimum extent at the northern and southern edges. 
We also explore the stability of N$_2$ ice deposits outside the latitudes and longitudes of the Sputnik Planitia basin. Results show that N$_2$ ice  is not stable at the poles but rather in the equatorial regions, in particular in depressions, where thick deposits may persist over tens of millions of years, before being trapped in Sputnik Planitia. 
Finally, another key result is that the minimum and maximum surface pressures obtained over the simulated millions of years remain in the range of milli-Pascals and Pascals, respectively. 
This suggests that Pluto never encountered conditions allowing liquid nitrogen to flow directly on its surface. 
Instead, we suggest that the numerous geomorphological evidences of past liquid flow observed on Pluto's surface are the result of liquid nitrogen that flowed at the base of thick ancient nitrogen glaciers, which have since disappeared.

\end{abstract}

\begin{keyword}
Pluto\sep nitrogen\sep paleo\sep Modelling\sep GCM\sep Sputnik~Planitia\sep\\
\texttt{http://icarus.cornell.edu/information/keywords.html}
\end{keyword}


\section{Introduction}
\label{secpaleo:intro}

\subsection{Pluto's ices observations}

Among the most striking observations of Pluto made by New Horizons in July 2015 is the prominent nitrogen ice sheet laying in Sputnik Planitia\footnote{The place names mentioned in this paper include a mix of officially approved names and informal names.} (SP), which displays a highly diverse range of terrains, as described by \citet{Whit:17,Moor:17,McKi:16}. First, bright nitrogen-rich plains (0$^{\circ}$-30$^{\circ}$N) contrast with darker plains at higher latitudes (30$^{\circ}$-40$^{\circ}$N) having higher amounts of diluted methane,  and even darker, more methane and tholins rich plains (40$^{\circ}$-50$^{\circ}$N) at the northern edge of SP (see composition maps Fig. 5.C in \citet{Prot:17} and Figs. 13.2, 15 and 18 in \citet{Schm:17}). Cellular patterns, indicative of active solid-state convection \citep{McKi:16,Trow:16}, are observed in the northern part of SP (0$^{\circ}$-40$^{\circ}$N) but not in the southern part of SP (0$^{\circ}$-25$^{\circ}$S). The absence of convection cells coincides with the presence of hundred meters deep pits on the surface of the ice sheet \citep{Moor:17}. Glacial flow activity is observed through the valleys at the eastern edge of SP (flowing toward the center of SP from the uplands, 20$^{\circ}$S-30$^{\circ}$N) and at the northern edge of SP (flowing outward the basin), as shown by \citet{Howa:17} and \citet{Umur:17}. Finally, rugged water ice mountains surround the SP region (Al-Idrisi, Hillary and Tenzing Montes), suggesting that they have been eroded and shaped over time by the action of glacial flow \citep{Ster:15,Howa:17}, as the terrestrial ``Nunataks'' in Greenland and Antarctica. 

New Horizons also revealed bright deposits of nitrogen outside SP. In the equatorial regions, the eastern side of Tombaugh Regio is covered by numerous patches of nitrogen ice, which are also observed further east in the depressions and valleys between the high altitude ``bladed terrains'' deposits \citep{Moor:17} and further west at the bottom of deep craters \citep{Schm:17}. In addition, bright patches of nitrogen ice are detected over a latitudinal band between 30$^{\circ}$N and 60$^{\circ}$N \citep{Schm:17}.

What is the history of Sputnik Planitia and the nitrogen deposits? 
Resurfacing by glacial flow, solid-state convection, or nitrogen sublimation and condensation have been proposed to explain the formation and disappearance of the pits and polygonal cells within SP \citep{Whit:17,Moor:17}. In addition, sublimation-condensation processes are thought to drive the difference in ice albedos, composition and distribution outside Sputnik Planitia \citep{Whit:17,Prot:17,Howa:17}. However, the timescales and amounts of ice involved are not known, which prevents us from distinguishing the roles of each process and the nature of the reservoirs (perennial/seasonal).

\subsection{Modelling of the present-day Pluto's volatile cycle}

The condensation and sublimation of Pluto's volatiles and their transport over several Pluto seasons has been modelled under current orbital conditions by \citet{BertForg:16}, that is with an obliquity of 119.6$^{\circ}$, an orbit eccentricity of 0.2488 and a solar longitude at perihelion of 3.8$^{\circ}$. Solar longitude at perihelion (L$_{s~peri}$) is the Pluto-Sun angle at perihelion, measured from the Northern Hemisphere spring equinox (L$_{s~peri}$=0$^{\circ}$ when the perihelion coincides with the northern spring equinox). It defines the link between Pluto season and the distance from the Sun (and thus also the duration of the season). 
Results using a seasonal thermal inertia for the sub-surface within 500-1500~J~s$^{-1/2}$~m$^{-2}$~K$^{-1}$ showed that nitrogen and carbon monoxide (CO) are trapped inside the Sputnik Planitia basin because its low altitude (its surface is located at $\sim$~3~km below the surroundings terrains) induces a higher condensation temperature of the ice (compared to the ice outside the basin), leading to an enhanced thermal infrared cooling and condensation rate at the bottom of the basin \citep{BertForg:16}. 
Note that the reverse occurs on Earth, where the ice caps form at the poles and at high altitude.
In these simulations, methane also accumulates in SP but its low volatility allows it to condense on warmer surfaces (where N$_2$ and CO would instantly sublimate) and form seasonal frosts of pure methane everywhere in the fall-winter hemisphere except at the equator which tends to remain free of volatile ice. 
Results also showed that bright methane frosts in the northern hemisphere could favour nitrogen condensation on it and thus lead to the formation of a seasonal nitrogen polar cap. These polar deposits sublime in spring from the pole and in 2015, only a latitudinal band of nitrogen frost around 45$^\circ$N remained. 

\subsection{The changes of obliquity and orbital parameters over astronomical timescales}

While \citet{BertForg:16} focused on the volatile cycles in Pluto's current orbital conditions, these latter are known to vary over timescales of 100~000 terrestrial years. Pluto's high obliquity varies between 104$^{\circ}$ and 127$^{\circ}$ over 2.8 Myrs. The solar longitude of perihelion of Pluto's orbit varies with a precession period of 3.7 Myrs, while its eccentricity oscillates between 0.222 and 0.266 with a 3.95 Myrs period  \citep{Earl:17,Binz:17,Hami:16}. 
The variation of these parameters, known as the Milankovitch mechanism, also occurs on the Earth, Mars and Titan. The parameters combine to modulate the solar insolation and surface temperatures, forcing the volatile ices that form glaciers, frost or lakes to migrate in different regions with time. 

On Earth, obliquity changes are known to have played a critical role in pacing glacial and interglacial eras.
Because of the absence of a big moon and its proximity to Jupiter, Mars has an obliquity which varies much more strongly than on Earth, and experienced periods when poles were warmer than equator, like on Pluto. 
Many surface structures of Mars are thought to be the effect of orbital forcing or of the Milankovitch cycle on the climate of Mars.
For instance, climate modeling efforts showed that during high obliquity periods, ice can be deposited almost anywhere in the mid-latitudes, explaining the evidences of glaciers and widespread ground-ice mantle in these regions, while during low obliquity ice is transported back to the poles \citep[e.g.]{Levr:07,Misc:03,Forg:06,Made:09}.
On Titan, the cooler summers in the north pole explain the lake’s preference for the northern latitudes \citep{Schn:12}. The differing solar insolation between both hemisphere would result from the eccentricity of Titan’s orbit and the obliquity of Saturn, coupled with Titan’s low inclination and obliquity \citep{Ahar:09}.
Climatic changes similar in scale to Earth’s climatic cycles are expected as the obliquity and orbital parameters of Titan vary on timescales of tens of thousands of years. 

By comparison with the Earth, Mars and Titan, Pluto's climate is expected to be dictated by the universal Milankovitch mechanism as well.
A few studies have explored the variation of insolation on Pluto caused by the changes of obliquity, solar longitude of perihelion and eccentricity, and have shown that the obliquity is the main driver of Pluto's insolation \citep{Earl:17,Binz:17,Ster:17}. 
However these thermal models neglected the impact of seasonal thermal inertia (TI), which strongly controls surface temperatures (see Section~\ref{secpaleo:surftemp}) and they did not address the transport of volatiles, necessary to fully understand how Pluto's ices evolved in the past. 

\subsection{Objectives of this paper}
\label{secpaleo:objectives}

Our objective is to investigate the evolution and distribution of nitrogen ice on Pluto over the past millions of Earth years (Myrs).
To do that, we extend the \citet{BertForg:16} study and use the Pluto's volatile transport model, taking into account (1) the changes of the astronomical cycles (obliquity, solar longitude of perihelion and eccentricity) induced by the perturbation of the Sun on the Pluto-Charon binary system \citep{Dobr:97}, (2) realistic reservoirs of nitrogen ice, (3) the changes of ice thickness induced by glacial flow. 

Pluto's astronomical cycles are thought to have been stable over at least the last 20 million years \citet{Binz:17}, and probably even before. The first reason is that Pluto's orbit is in a relatively isolated region of the Solar System, never getting within $\sim$11~AU of any major planet, and it is therefore subject to very little perturbations \citep{Dobr:97}. In addition, \citet{Binz:17} state that the presence of ancient craters at the equator demonstrates a certain stability of the astronomical cycles, which could extend back in time by hundreds of Myrs (otherwise the craters would have been eroded away or completely buried). In this paper, we assume that the astronomical cycles remained stable during the last 30 Myrs.

Here we explore the impact of orbital and obliquity changes on the nitrogen cycle only. Thus, all the simulations of this paper are performed without the cycles of methane and CO. This choice is driven by the fact that the number of sensitivity parameters and initial states explored in this work already makes it start at a certain complexity level. In fact, we know from \citet{BertForg:16} that CO ``follows`` N$_2$ and remains always trapped in N$_2$ ice. The case of CH$_4$ ice is much more complex because it can form CH$_4$-rich deposits (with 3-5$\%$ N$_2$) on Pluto's surface and trigger N$_2$ condensation if its albedo is high enough. Exploring the cycle of methane over astronomical timescales and its impact on Pluto's climate will be the topic of a separate paper. 

In Section~\ref{secpaleo:model}, we describe the Pluto volatile transport model and its recent development allowing for the simulations of the nitrogen cycle over several astronomical cycles (in particular, the model includes an ice redistribution algorithm, glacial flow modelling, changes of topography, obliquity, and orbital parameters with time). 

In Section~\ref{secpaleo:surftemp}, we show the impact of the obliquity, the orbital parameters and the thermal inertia on the surface temperatures averaged over the past Myrs.  

Then, in Section~\ref{secpaleo:SP}, we investigate how the past cycles of nitrogen sublimation and condensation (at diurnal, seasonal, astronomical timescales), as well as the glacial flow of N$_2$ ice may have shaped Sputnik Planitia as it is observed today. 

In Section~\ref{secpaleo:eqstates}, we explore possible steady state for nitrogen deposits outside Sputnik Planitia, by performing simulations over the last 30 Myrs with different reservoirs of N$_2$ ice initially placed at the poles, at the equator or uniformly spread over the surface. We also explore the maximum and minimum surface pressures Pluto encountered during that time. We discuss these results in Section~\ref{secpaleo:discussion}.

\section{Model description}
\label{secpaleo:model}

We use the latest version of the Pluto volatile transport model of the Laboratoire de M\'et\'eorologie Dynamique (LMD) \citep[see Methods]{BertForg:16}.

This model represents the physical processes that control the condensation and sublimation of Pluto's volatiles (insolation, suface thermal balance) and uses a simple global mixing function to parametrize the atmopsheric transport and dynamics. 
Note that in this model, the atmosphere is considered transparent: there is no atmospheric process taken into account aside from the condensation, sublimation and exchanges of latent heat with the surface. Such a model works well on Pluto because the surface energy balance is not significantly sensitive to the atmospheric sensible heat flux and to the radiative transfer through the air.


In this section, we describe the grid and surface properties used in our simulations (Sections~\ref{secpaleo:grid} and \ref{secpaleo:surfprop}) and the recent developments performed in the code, allowing the simulation of the nitrogen cycles over astronomical timescales. These improvements concern a paleoclimate mode (Section~\ref{secpaleo:paleomode}), the implementation of the latest topography of Pluto with a specific relief for the SP basin (Section~\ref{secpaleo:topo}) and a glacial flow scheme (Section~\ref{secpaleo:flow}).
Note that all figures and maps are shown using the IAU convention, spin north system for definition of the north pole \citep{Buie:97,Zang:15}, that is with spring-summer in the northern hemisphere during the 21th Century. 

\subsection{Model grid}
\label{secpaleo:grid}

In this paper, the simulations investigating the stability of N$_2$ ice deposits outside SP (Section~\ref{secpaleo:eqstates}) have been performed with a horizontal grid of 32$\times$24 points, that is a grid-point spacing of 7.5$^{\circ}$ latitude by 11.25$^{\circ}$ longitude and a spatial resolution of about 150~km. Simulations focusing on the N$_2$ cycle within Sputnik Planitia (Section~\ref{secpaleo:SP}) have been performed with a twice higher spatial resolution of 5.6$^{\circ}$ in longitude and 1.875$^{\circ}$ in latitude.

\subsection{Model grid and surface properties}
\label{secpaleo:surfprop}

As in \citet{BertForg:16}, the reference albedo of nitrogen ice is set to 0.7, and its emissivity to 0.8. The albedo and emissivity of the bare ground (free of N$_2$ ice) are set to 0.1 and 1 respectively. 
In the soil model, the near-surface layers have a low-TI to capture the short-period diurnal thermal waves, while the deeper layers have a high-TI to capture the much longer seasonal thermal waves. The diurnal TI is set to 20~J~s$^{-1/2}$~m$^{-2}$K$^{-1}$ (or SI), as inferred from Spitzer thermal observations \citep{Lell:11b}. In this paper, simulations have been performed without a diurnal cycle (the insolation is averaged over the Pluto day) and therefore the diurnal TI has no impact on the results. The reference seasonal thermal inertia TI is uniformly set to 800~SI, because it corresponds to the best case simulation reproducing the threefold increase of surface pressure observed between 1988 and 2015 \citep{BertForg:16} and the 1-1.2 Pa value in 2015. Here we assume that all terrains (water ice bedrock and nitrogen ice) have the same TI. In this paper, sensitivity simulations have also been performed using 400 and 1200~SI. The modelled diurnal and annual skin depths are 0.008~m and 20-60~m respectively. 

To adequately resolve these scale lengths, the subsurface is divided into 24 discrete layers, with a geometrically stretched distribution of layers with higher resolution near the surface to capture the short waves (the depth of the first layer is z$_1$=1.4$\times$10$^{-4}$m) and a coarser grid for deeper layers and long waves (the deepest layer depth is near 1000~m). 
The depth of each layer is given by: 
\begin{equation}
 z_k = z_1 2^{k-1}
\end{equation}

Our simulations are performed assuming no internal heat flux. Adding an internal heat flux of few mW~m$^{-2}$, as suggested in \citet{RobuNimm:11} for Pluto, does not change significantly the results. Our tests show that the surface temperature increases by 0.2~K when taking into account an internal heat flux of 3~mW~m$^{-2}$ (see the discussion in \ref{secpaleo:flow}). 

Note that the seasonal thermal inertia of N$_2$ ice does not impact the amount of condensed or sublimed N$_2$ ice to first order.

Indeed, as shown by equation 12 and 13 in \citet{Forg:17}, the variation of the exchanged mass between the surface and the atmosphere $\delta$m$_0$ is nearly proportional to a product involving the surface heat capacity $c_s$ (in J~m$^{-2}$~K$^{-1}$), which depends on the thermal inertia:

\begin{equation}
 \delta m_0 \propto \frac{c_s}{L_{\mbox{\scriptsize N$_2$}}}~\Delta T_s
\end{equation}

with $L_{\mbox{\scriptsize N$_2$}}$ the latent heat of N$_2$ ($2.5~10^5$~J~kg$^{-1}$) and T$_s$ the surface temperature. Yet $\Delta$T$_s$ is nearly proportional to $\frac{F}{c_s}$ to first order, with $F$ the thermal flux absorbed by the surface (W~m$^{-2}$). Hence:

\begin{equation}
 \delta m_0 \propto \frac{F}{L_{\mbox{\scriptsize N$_2$}}}
\end{equation}

Consequently, to first order, $\delta$m$_0$ is independent of c$_s$ and of thermal inertia. 

\subsection{The paleoclimate mode and ice equilibration algorithm}
\label{secpaleo:paleomode}

Because the solar flux received by Pluto's surface is very low (about 1~W~m$^{-2}$), and because Pluto makes a full orbit around the Sun every 248 years, the modelled surface and subsurface temperatures and the surface ice distribution require simulations of several Pluto years to reach a steady state. Running the Pluto volatile transport model with the N$_2$ cycle only can take around 5 minutes of computing time for one simulated Pluto year at the chosen resolution. Consequently, running climate evolution over 1 Myrs would require 2 weeks of simulation and performing paleoclimate simulations over several astronomical cycles, e.g. over the last 30 Myrs, would be prohibitively time-consuming.
To resolve this problem, we implemented in the model a paleoclimate mode containing an ice iteration scheme. 
The algorithm is similar to the approach taken for Mars simulations in \citet{Word:13}:
Starting from the initial surface ice distribution $q^-$ (kg~m$^{-2}$), the model is ran normally for 5 Pluto years (Step 1), in order to reach repeatable annual condensation and sublimation rates as well as repeatable annual temperatures variations. In the last year, the annual mean ice rate of change $<dq/dt>$ is evaluated at each grid point and then multiplied by a multiple-year timestep $\Delta$t to give the updated surface ice distribution $q^+$ and reservoir (Step 2):

\begin{equation}
q^+_{ice}=q^-_{ice}+<\frac{dq_{ice}}{dt}>\Delta t
\end{equation}

If the reservoir of ice at one gridpoint has entirely sublimed during the “paleo-timestep” $\Delta$t, the amount of ice is set to 0~kg~m$^{-2}$, and after redistribution, the amount in each cell is normalized to conserve the total nitrogen mass (ice and vapour) in the system. 
Then, the topography is updated according to the new thickness of the deposits on the surface (Step 3).
Finally, the orbital parameters and the obliquity are changed according to the new date of the simulation (Step 4).  
Then the loop started again: the model is run again for 5 Pluto years ; at each paleo-timestep $\Delta$t, the topography is updated according to the annual mean ice rate of the last Pluto year, and the orbital parameters and the obliquity of Pluto are changed according to the new date t+$\Delta$t. 

The paleo-timestep must be small enough so that the changes of obliquity and orbital parameters allow the surface ice distribution to reach a steady state at each paleo timestep, but it must be high enough to reduce significantly the computing time of the simulation. In our simulations described here, we typically used $\Delta$t = 50~000 Earth years, which corresponds to about 200 Pluto orbits and a maximal change in its obliquity of 1$^{\circ}$ of latitude \citep{Binz:17}.

\begin{figure}[!h]
\begin{center} 
	\includegraphics[width=15cm]{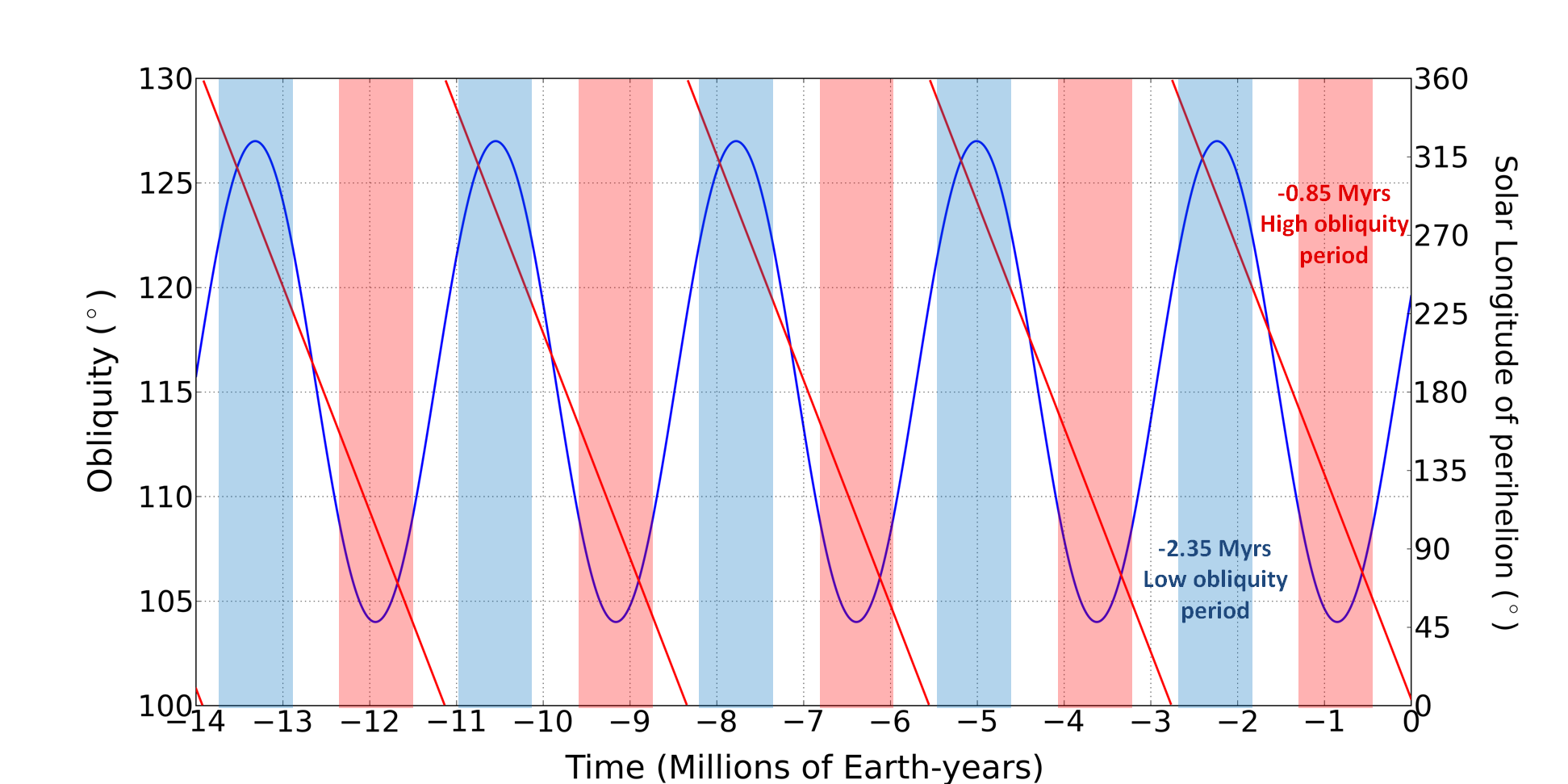}
\end{center} 
\caption{The astronomical cycles of Pluto during the past 14 Earth million years ($\sim$ 5 obliquity cycles): cycles of obliquity (blue solid line, period: 2.8 Myrs) and L$_{s~peri}$ (red solid line, period: 3.7 Myrs).
High obliquity refers to values close to 90$^\circ$, while ''low`` obliquity is used here to designate the periods with minimal obliquity, that is far from 90$^\circ$, although here it remains relatively high compared to other bodies in the Solar System.
Present-day Pluto (obliquity=119.6$^{\circ}$, that is 60.4$^\circ$ in retrograde rotation) is subject to an intermediate obliquity (white background). A low obliquity period occurred 2.35 Myrs ago (blue background) and a high obliquity period 0.9 Myrs ago (red background).} 
\label{orbitchanges}
\end{figure}

The changes of obliquity, solar longitude of perihelion and eccentricity are taken from \citet{Earl:17} and \citet{Dobr:97} and are shown on \autoref{orbitchanges}. 
Assuming that the obliquity and orbital parameters remain periodic with time, we extrapolated the data back to 30 Myrs ago, which corresponds to the starting date of our simulations. Here the variations of eccentricity (between e$_{min}$=0.222 and e$_{max}$=0.266) are taken into account but not shown for the sake of simplicity. Note that its impact on Pluto's climate is negligible compared to the change of the two other parameters \citep{Hami:16}.

\subsection{Topography}
\label{secpaleo:topo}

The simulations are performed with the latest topography data of Pluto \citep{Sche:18}. In the south hemisphere, where there is no data, we considered a flat surface (at mean radius). If we model a surface topography in the southern hemisphere varying of few kilometers, results remain unchanged and N$_2$ ice will accumulate at similar latitudes. 

In the simulations performed with glacial flow of N$_2$ ice, we modified the topography of Sputnik Planitia by placing the bedrock much deeper than the actual surface of SP, in order to represent SP with realistic amounts of ice. We assume that the bedrock below the center of SP is a 10~km deep elliptical basin, which is in the range of the estimates for the thickness of volatiles where polygonal cells are observed \citep{Moor:16, McKi:16, Trow:16, Kean:16}. 

\begin{figure}[!h]
\begin{center} 
	\includegraphics[width=14cm]{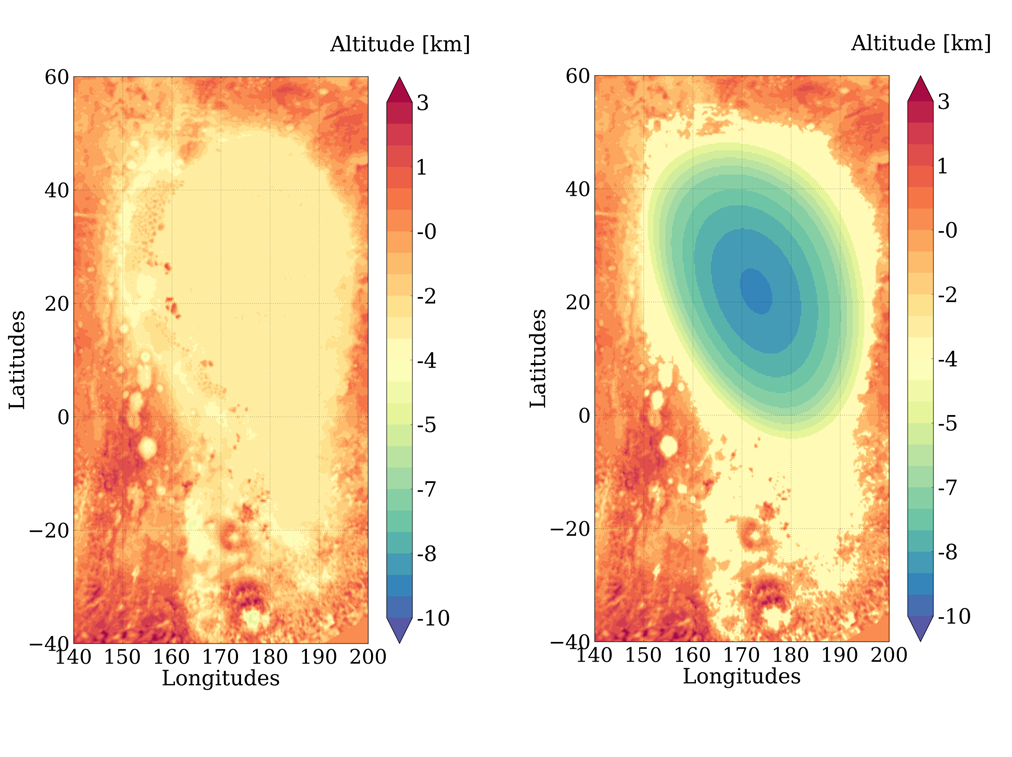}
\end{center} 
\caption{Left: Topography of Sputnik Planitia as seen by New Horizons, filled by ice at 2.5~km below the mean surface \citep{Sche:16AGU,Sche:16LPSC}. Right: The modeled initial topography of Sputnik Planitia with a 10~km deep bedrock (here not filled by nitrogen ice) assumed in the model.\newline} 
\label{topopluton}
\end{figure}

The elliptical basin covers the latitudes 10$^{\circ}$S-50$^{\circ}$N (\autoref{topopluton}), with a semi major axis of 1200~km and the foci F=(42$^{\circ}$N,163$^{\circ}$E) and F'=(1.75$^{\circ}$N,177$^{\circ}$W).
The edges and the southern parts of the basin are less deep, in accordance with typical impact basin shapes and the absence of convective cells there. We have different cases where we placed the bedrock in these areas at 4~km or 5~km below the mean surface level, as shown by \autoref{topopluton}. In the simulations focusing on N$_2$ ice inside SP (Section~\ref{secpaleo:SP}), this modelled basin is then filled with N$_2$ ice up to 2.5~km below the mean surface level, which corresponds to the observed altitude of the surface of the SP ice sheet (in that case, the thickness of ice at the centre of SP would be 10-2.5=7.5~km). Note that in the modeled topography we removed the water ice mountains from Al-Idrisi to Hillary Montes for the sake of simplicity. 

\subsection{Ice viscous flow modelling}
\label{secpaleo:flow}

Despite being completely frozen, Pluto’s surface is remarkably active. Evidence of current and past flowing nitrogen ice have been observed by New Horizons all around Sputnik Planitia (rugged eroded water ice mountains, glaciers with moraine-like ridges...) and at higher latitudes \citep{Ster:17,Howa:17,Umur:17}.

In order to represent the nitrogen ice flow in the model, we use a laminar glacial flow scheme, as presented in \citet{Umur:17}, which is based on the N$_2$ ice rheology for low surface temperatures described in \citet{Yama:10} and depends on the thickness and temperature of the ice. 
The model has several limitations. First, very little laboratory data is available under Plutonian surface conditions and the rheology of solid N$_2$ has not been very well constrained to date (see \citet{Umur:17}), not to mention the rheology of the possible ice mixtures on Pluto and of the $\alpha$ crystalline structure. N$_2$ ice behaves as a viscous fluid with a viscosity ranging between 10$^8$-10$^{12}$ Pa~s at stresses of 10$^5$ Pa \citep{Durh:10}, and flows more rapidly than water ice on Earth, despite the lower gravity of Pluto. 
Secondly, the model strongly depends on the height of the column of ice, its temperature and the N$_2$ ice rheology properties, which are poorly known \citep{Umur:17}. In addition, we are limited by the horizontal resolution, which prevents us to reproduce with precision small glacial flows (e.g. in the narrow channels east of SP or around Tenzing Montes in the southern part of SP). 
Consequently, the model of glacial flow used in this paper does not intend to be quantitatively accurate given its simplicity and the unknowns 
but instead intends to reproduce to first order the activity of the nitrogen glaciers (e.g. in Sputnik Planitia rapid flows in the centre of Sputnik Planitia and slow flows on the edges). 

We use the scheme described in \citet{Umur:17} under the following assumptions. First, we consider that the ice within SP flows like pure nitrogen ice. Then, we consider the simple case of a laminar flow with an isothermal ice, without basal melting (thus we are in the case of a “basally cold and dry” glacier). This is typically valid for thin layers of ice (shallow ice-sheet approximation). In fact, a conductive temperature gradient of 15~K~km$^{-1}$ due to internal heat flux on Pluto is suggested in \citep{McKi:16}, assuming no convection. In that case, basal melting would occur below about 2~km of N$_2$ ice, assuming a surface temperature of 37~K.
This 15~K~km$^{-1}$ value is probably an upper limit because of the convection within the ice. \citep{McKi:16} suggest that the ice layer is convecting in the so-called sluggish lid regime, which involves the entire layer of ice in the overturn. Therefore the ice temperature within the layer is likely colder than the temperature assumed without convection.
\citep{McKi:16} use a Nusselt number of 3.2, which means that the effective temperature gradient (in the horizontal mean) is rather 5~K~km$^{-1}$. In that case, basal melting would occur below about 5~km of N$_2$ ice. 

If the surface temperature of the ice has approached the melting temperature of 63~K at the triple-point of N$_2$ in the past, as suggested in \citep{Ster:17}, then it is likely that the thin layers of ice at the edges of the ice sheet were ``temperate'' glaciers at this time. 
However, our results show that during an entire astronomical cycle (even during high obliquity periods), the surface temperatures of nitrogen ice remain low, below 40~K (see Section~\ref{secpaleo:pressures}), which reinforces our assumption of dry glacier at the edges of SP. 
Here we apply the case of a “basally cold and dry” glacier to all encountered ice thicknesses (with no internal heat flux and no basal melting).
This is acceptable to first order because (1) our study focus on the edges of SP and on the glaciers outside SP, whose thickness in the model remains thin ($<$1~km), (2) the large amount of ice in the centre of SP already flows extremely rapidly; a basal melting here would allow for even more rapid flow which would not significantly impact our results, (3) although the impact of internal heat flux on soil and surface temperature is not negligible (+0.2~K at the surface), it has a small effect on the flow compared to other parameters of the model, which are not well constrained, such as the albedo and the thickness of the ice (the glacial flow modelling strongly depends on the depth of the bedrock). 
We tested the model assuming that the effective temperature controlling the glacial flow is the one that we would obtain at the bottom of the glacier taking into account a conductive temperature gradient of 15~K~km$^{-1}$ (that is $\sim$55~K for 1~km thick glacier), and it does not change significantly the results of this paper.

Finally, we consider that the bedrock remains static and is not altered by the glacial flow. 
Consequently, in the model, the ice is transferred from one grid point to another one using the modified Arrhenius-Glen analytic function of the mass-flux given in \citet{Umur:17}:

\begin{equation}
q_{0}=g_Q exp[\frac{\frac{H}{H_a}}{1+\frac{H}{H_{\Delta T}}} ] q_{glen} 
\end{equation}
\begin{equation}
q_{glen}=A(\rho g)^{n}\frac{H^{n+2}}{n+2}\frac{tan^{n-1}(\theta) } { (1+tan^{2}(\theta) ) ^{\frac{n}{2}}}
\end{equation}

With q$_0$ in m$^2$~s$^{-1}$, $g$ the gravity at Pluto's surface (0.6192~m~s$^{-1}$), $H$ the ice thickness of the considered column of ice. 
$\rho$ is the nitrogen ice density, set to 1000~kg~m$^{-3}$ \citep{Scot:76,McKi:16,Umur:17}. 
g$_Q$ is a corrective factor given in \citet{Umur:17} and set to 0.5. 
H$_a$ and H$_{\Delta T}$ are parameters depending on the surface pressure and are given by equation 14 in \citet{Umur:17}. 
$\theta$ is the angle between the two adjacent columns of ice (see Figure 7 in \citet{Umur:17}), and is defined by $\theta$=arctan($\Delta$H/L$_{ref}$), with $\Delta$H the difference of altitude between both columns (computed from the bedrock topography and the amount of ice) and L$_{ref}$ is the characteristic distance of the glacial flow (distance between both columns, that is both adjacent grid-points). The parameters $A$ and $n$ are given in \citet{Umur:17} are only depend on the surface temperature. 
By using this scheme in our model, we obtain the same relaxation times for the ice than those shown on Figure 8-9 in \citet{Umur:17} (the corresponding relaxation time of a 50~km long channel, sloping at $\theta$= 10$^{\circ}$ and initiated with 200 m of glacial ice is about 50 years).
We adapted this scheme so that it fits on the model grid and so that each grid point can redistribute the correct amount of ice to the neighboring points.

\section{Orbital, obliquity and TI changes as drivers of surface temperatures}
\label{secpaleo:surftemp}
Obliquity is the main driver of insolation changes on Pluto. The polar regions receive more flux than the equator on annual average during high obliquity periods and about the same flux than the equator during low obliquity periods. 
However, although one could think the rule also applies for surface temperatures, it is not systematically the case. The following paragraph explains why.
 
If one assumes that the surface temperature is only driven by the absorbed flux and the infrared cooling (neglecting the soil TI, the latent and sensible heat flux…), then the surface temperature at equilibrium is given by:

\begin{equation}
T_{eq} = \sqrt[4]{\frac{(1-A)F}{\epsilon~\sigma}}
\end{equation}

with $A$ and $\epsilon$ the surface albedo and emissivity respectively, $F$ the absorbed flux and $\sigma$ the Boltzmann constant. 
This equation shows that when the flux $F$ strongly increases, the surface temperature only increases by a factor of F$^{1/4}$. In other words, the thermal infrared cooling limits the increase of surface temperatures.

\begin{figure}[!h]
\begin{center} 
	\includegraphics[width=15cm]{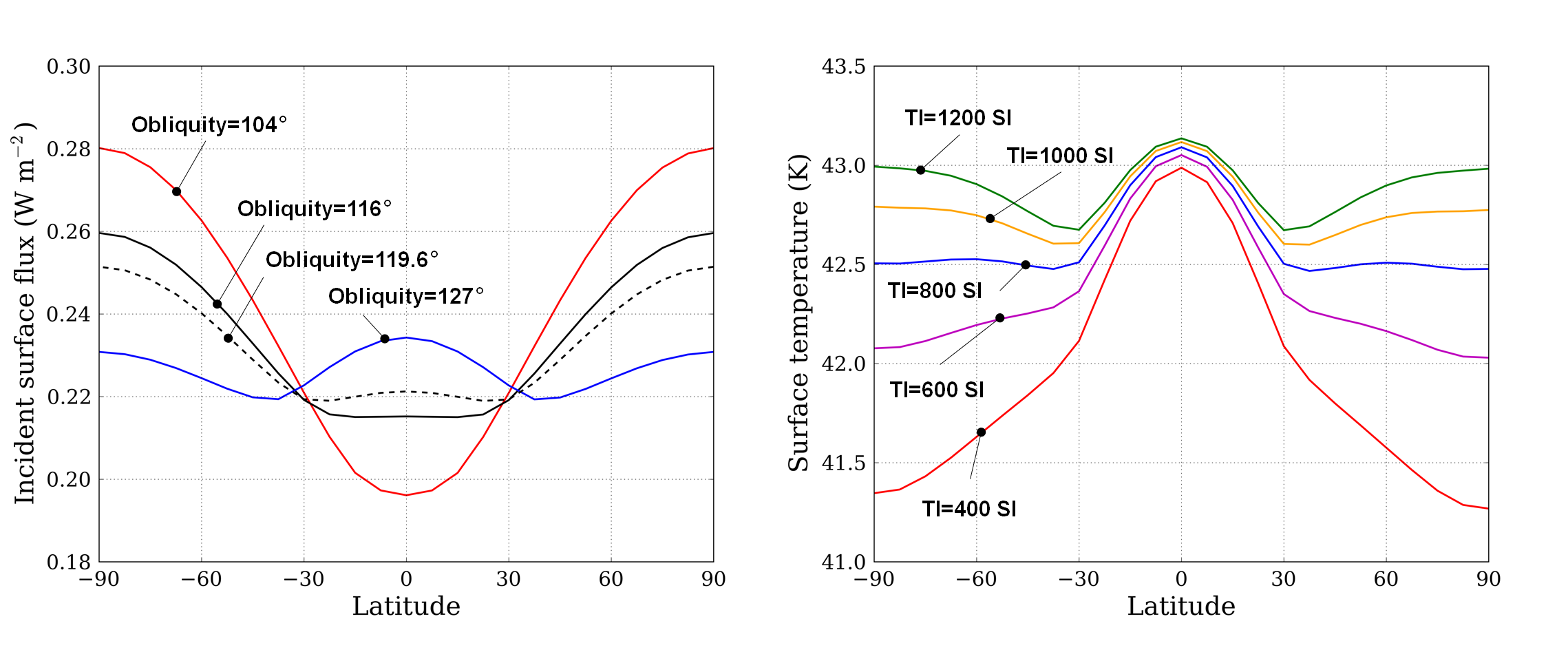}
\end{center} 
\caption{Surface thermal model results. Left: Annual mean incident solar flux for an obliquity of 104$^{\circ}$, 116$^{\circ}$, 119.6$^{\circ}$ and 127$^{\circ}$. Generally speaking, the poles receive more flux in average than the equatorial regions, except during the low obliquity periods (127$^{\circ}$) where mid-latitudes receive less flux in average. Right: Annual mean surface temperatures obtained with the obliquity of 119.5$^{\circ}$ and the incident solar flux shown on the left panel, a L$_{s~peri}$ of 0$^{\circ}$ and TI of 400 SI (red), 600 SI (purple), 800 SI (blue), 1000 SI (orange), 1200 SI (green). The surface albedo is uniformly set to 0.1, and the emissivity to 1. The coldest points are the poles for the low TI case and the ``low latitudes bands'' at $\pm$ 30$^{\circ}$ for the high TI case.\newline} 
\label{fluxtemp}
\end{figure}

As a result, the surface temperatures at the poles do not increase as much as the insolation during summer, and the poles can be colder than the equator on annual average, even if the mean insolation is not. This is true for low thermal inertia ($<$~800 SI), but not for medium-to-high TI, which enables the poles to store the heat accumulated during summer and release it during winter, as illustrated on \autoref{fluxtemp}. In the cases of TI between 800-1200~SI, the poles and the equator have similar surface temperatures and the coldest regions are around $\pm$30$^{\circ}$ latitude. The higher the TI, the more equatorial are the coldest regions.  

\begin{figure}[!h]
\begin{center} 
	\includegraphics[width=15cm]{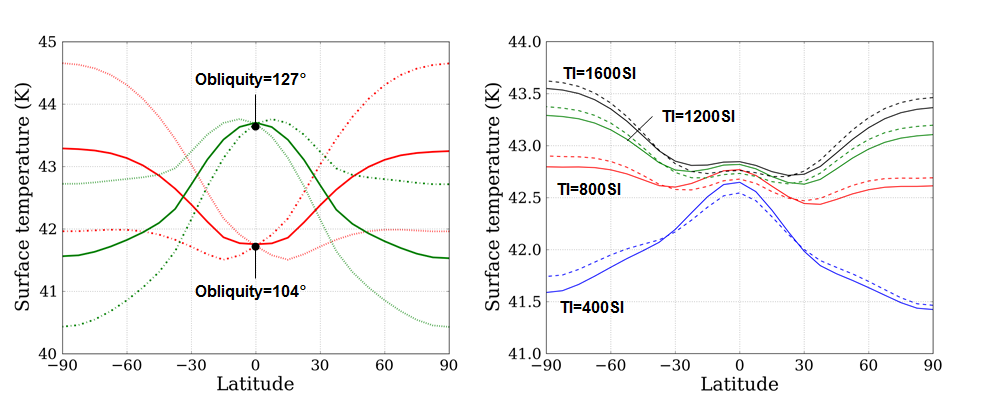}
\end{center} 
\caption{Surface thermal model results. Left: Annual mean surface temperatures obtained assuming uniform and constant surface conditions (surface albedo = 0.1, TI=800 SI) for the maximum and minimum values of obliquities (red: 104$^{\circ}$, green:127$^{\circ}$) and L$_{s~peri}$ (solid line: 0$^{\circ}$, dashed: 90$^{\circ}$, dot-dashed: 270$^{\circ}$). While the obliquity and the TI drive the location of the coldest region (polar or equatorial regions), the L$_{s~peri}$ induces an asymmetry of temperatures with Ls=90$^{\circ}$ and Ls=270$^{\circ}$ leading to a colder and warmer north pole respectively. Right: Surface temperatures averaged over the last 14 Myrs (last 5 obliquity cycles, solid lines) and 2.8 Myrs (last obliquity cycle, dashed lines) for different cases of TI. The surface albedo is uniformly set to 0.1, and the emissivity to 1. The polar regions are warmer than the equatorial regions, except in the case of TI lower than 800 SI. Our reference simulation (TI=800 SI) shows that the regions around $\pm$ 30$^{\circ}$ are colder in average.\newline} 
\label{tempmean}
\end{figure} 

The variation of L$_{s~peri}$ is also significant because of the high eccentricity of Pluto's orbit. In the past Myrs, the L$_{s~peri}$ parameter has created a North-South asymmetry of annual mean insolation and surface temperatures, favouring a warmer southern hemisphere when L$_{s~peri}$ values were close to 90$^{\circ}$ and a warmer northern hemisphere when L$_{s~peri}$ values were close to 270$^{\circ}$ (\autoref{tempmean}.A). The surface temperatures tend to be the same in both hemispheres for L$_{s~peri}$ values close to 0$^{\circ}$ and 180$^{\circ}$.

The surface temperatures averaged over the last 14 Myrs (which corresponds to the last 5 obliquity cycles) and over the last 2.8 Myrs are shown on \autoref{tempmean}.B. For the same reasons mentioned above, the equatorial regions are colder than the poles for medium-to-high TI and warmer for low TI ($<$~800 SI). For medium-to-high TI, the lowest temperatures are obtained at 30$^\circ$N-45$^\circ$N, which corresponds to the latitudes where a band of nitrogen ice has been observed by New Horizons. It can also be noted that the northern hemisphere is in average over several Myrs sligthly colder than the southern hemisphere in all TI cases. This is because the last high obliquity periods of Pluto's past (during which the poles receive the most of insolation) remained coupled with a solar longitude at perihelion close to 90$^{\circ}$, thus favouring colder northern latitudes during these periods and in average over several Myrs. 

\vspace{1cm}

\begin{figure}[!h]
\begin{center} 
	\includegraphics[width=15cm]{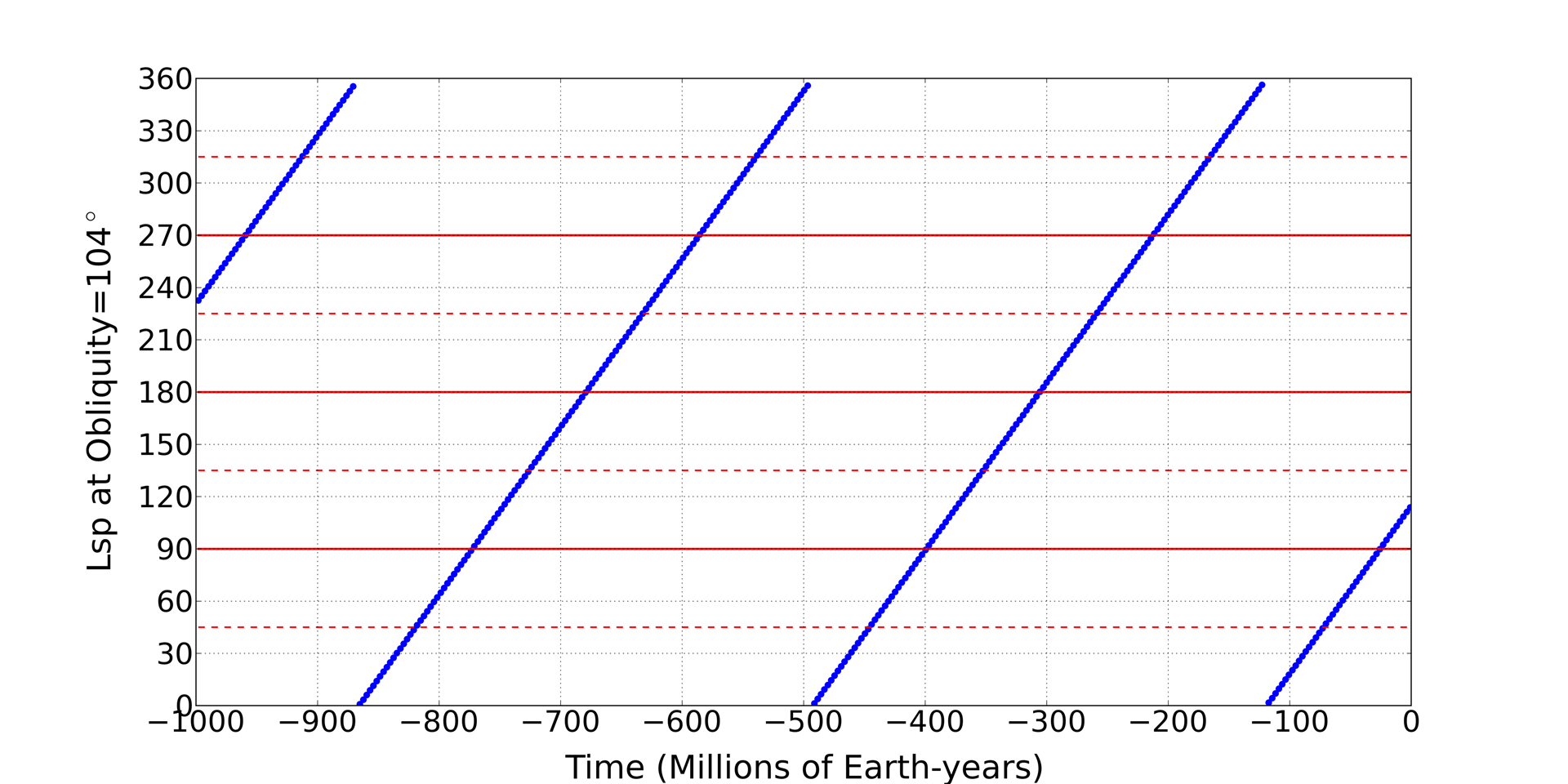}
\end{center} 
\caption{Evolution of the solar longitude of perihelion (L$_{s~peri}$) at high obliquity (104$^{\circ}$) during the last 1000 Myrs (assuming astronomical cycles stable with time). The L$_{s~peri}$ at high obliquity has been slowly shifted with time, e.g. from 110.8 to 113.4$^{\circ}$ during the last 6 Myrs, due to the sligth difference of periods between both L$_{s~peri}$ and obliquity cycles. During the last 70 Myrs, the L$_{s~peri}$ value at high obliquity remained close to 90$^{\circ}$ and thus lead to an asymmetry of insolation and surface temperatures favouring a sligthly warmer south hemisphere (see \autoref{tempmean}).\newline} 

\label{lspevol}
\end{figure}

Assuming that the evolution of obliquity and L$_{s~peri}$ remained stable during the last billion of years, one can quantify the shift between the obliquity and the L$_{s~peri}$ values. As shown by \autoref{lspevol}, between 260 and 165 Myrs ago, the L$_{s~peri}$ during high obliquity periods varied from 225$^{\circ}$ to 315$^{\circ}$, which favoured colder southern latitudes in average over several Myrs. Between 165 and 70 Myrs ago, the L$_{s~peri}$ during high obliquity periods varied from -45$^{\circ}$ to +45$^{\circ}$, leading to symmetric surface temperatures between both hemisphere in average over several Myrs. Finally, from 70 Myrs up to now, the L$_{s~peri}$ during high obliquity periods varied from 45$^{\circ}$ to 113$^{\circ}$, which favoured colder northern latitudes in average over several Myrs. The entire period of the cycle obliquity+L$_{s~peri}$ is 375 Myrs.

\section{Exploring the changes of N$_2$ ice thickness in Sputnik Planitia}
\label{secpaleo:SP}

In this section, we explore the past evolution of the N$_2$ ice thickness within Sputnik Planitia using the volatile transport model in the configuration as described above and with all of the initial N$_2$ ice reservoir sequestered in the deep Sputnik Planitia basin. We explore the changes of N$_2$ ice thickness considering its condensation and sublimation cycles, first without glacial flow (Section~\ref{secpaleo:cycles}) and then with glacial flow (Section~\ref{secpaleo:cyclesflow}). In this paper we assume a compact N$_2$-rich ice so that 1~kg of ice per m$^2$ corresponds to a thickness of 1~mm.

\subsection{The cycles of condensation and sublimation}
\label{secpaleo:cycles}

We first start the simulation 30 Myrs ago with nitrogen ice sequestered in SP and let the amount of ice evolve at the surface. 
In this section, we do not perform the simulations with glacial flow. Instead, we assume that the timescale for ice viscous flow is very short and that the ice sheet surface is effectively a level sheet, remaining flat at all times, and we only evaluate the condensation and sublimation rates within SP.
Note that in this simulation, no nitrogen frost form outside SP. 

\begin{figure}[!h]
\begin{center} 
	\includegraphics[width=16cm]{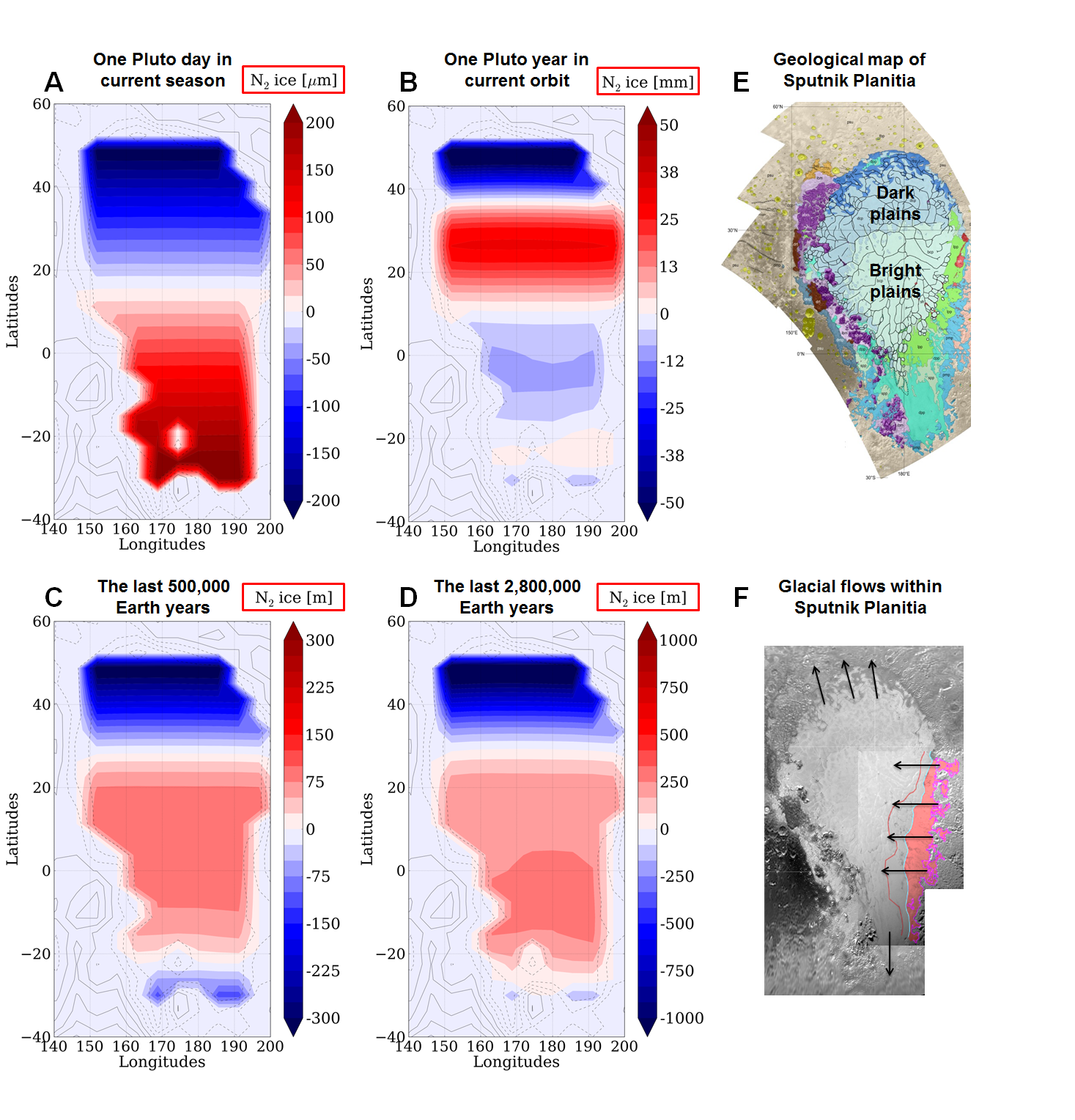}
\end{center} 
\caption{Sublimation-condensation rates of N$_2$ (zoom at Sputnik Planitia) at different timescales (note the order of magnitude differences between the colorbars, from $\mu$m to m): (A) During one Pluto day in July 2015. (B) During one Pluto year in current orbital conditions. (C) During the last 0.5 Myrs. (D) During the last 2.8 Myrs (obliquity cycle). (E) Geological map of the Sputnik Planitia region (a full resolution version can be found in \citet{Whit:17}. (F) New Horizons mosaic of Sputnik Planitia, with recent glacial activity indicated by the red area. The purple line indicates the extent of the N$_2$ ice sourced for the glaciation, the cyan line indicates the current ice deposition limit, and the red line indicates the inferred former ice deposition limit. The black arrows indicate the direction of the flow. Originally shown as Fig. 6 in \citet{Howa:17}.\newline} 
\label{maps}
\end{figure}

\autoref{maps} shows the net change of N$_2$ ice thickness obtained with the model over four different timescales: A, one Pluto day in July 2015; B, one current Pluto year; C, during the last 500~000 Earth years, which correspond to the estimated time of full resurfacing of SP by the action of the convection cells \citep{McKi:16}; D, during the last 2.8 Myrs (last obliquity cycle). These results are compared with geologic features observed by New Horizons within SP (\autoref{maps}.E-F). 

\clearpage

\begin{figure}[!h]
\begin{center} 
	\includegraphics[width=16cm]{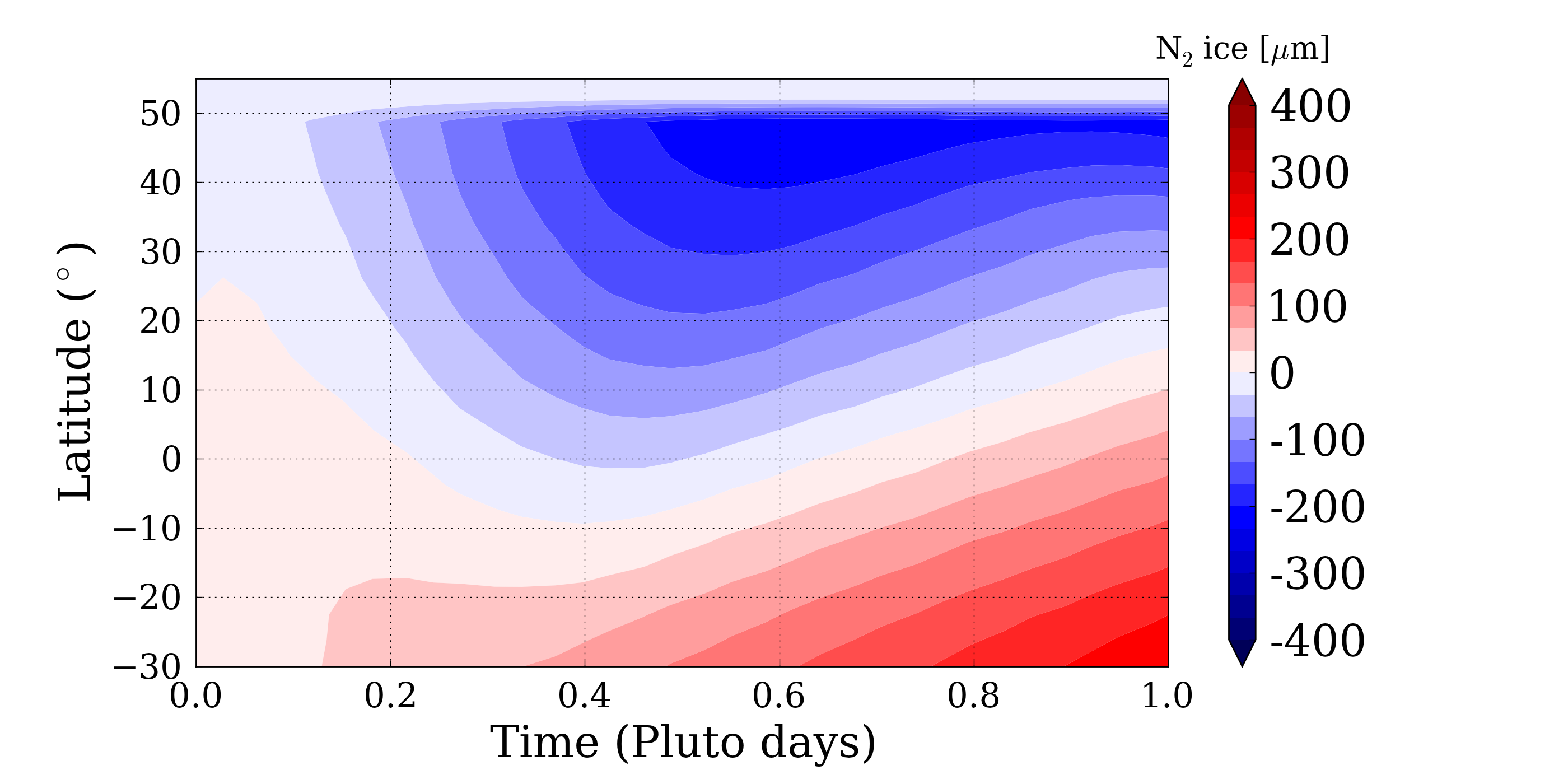}
\end{center} 
\caption{Variations of N$_2$ ice thickness within Sputnik Planitia, during one Pluto day in July 2015, normalized to 0 at t=0. The data is taken at the longitude 180$^{\circ}$, where the ice covers the latitudes 30$^{\circ}$S-50$^{\circ}$N. As shown by \autoref{maps}, the flux does not vary with longitude within Sputnik Planitia.} 
\label{evoldiurn_brut}
\end{figure}

\begin{figure}[!h]
\begin{center} 
	\includegraphics[width=16cm]{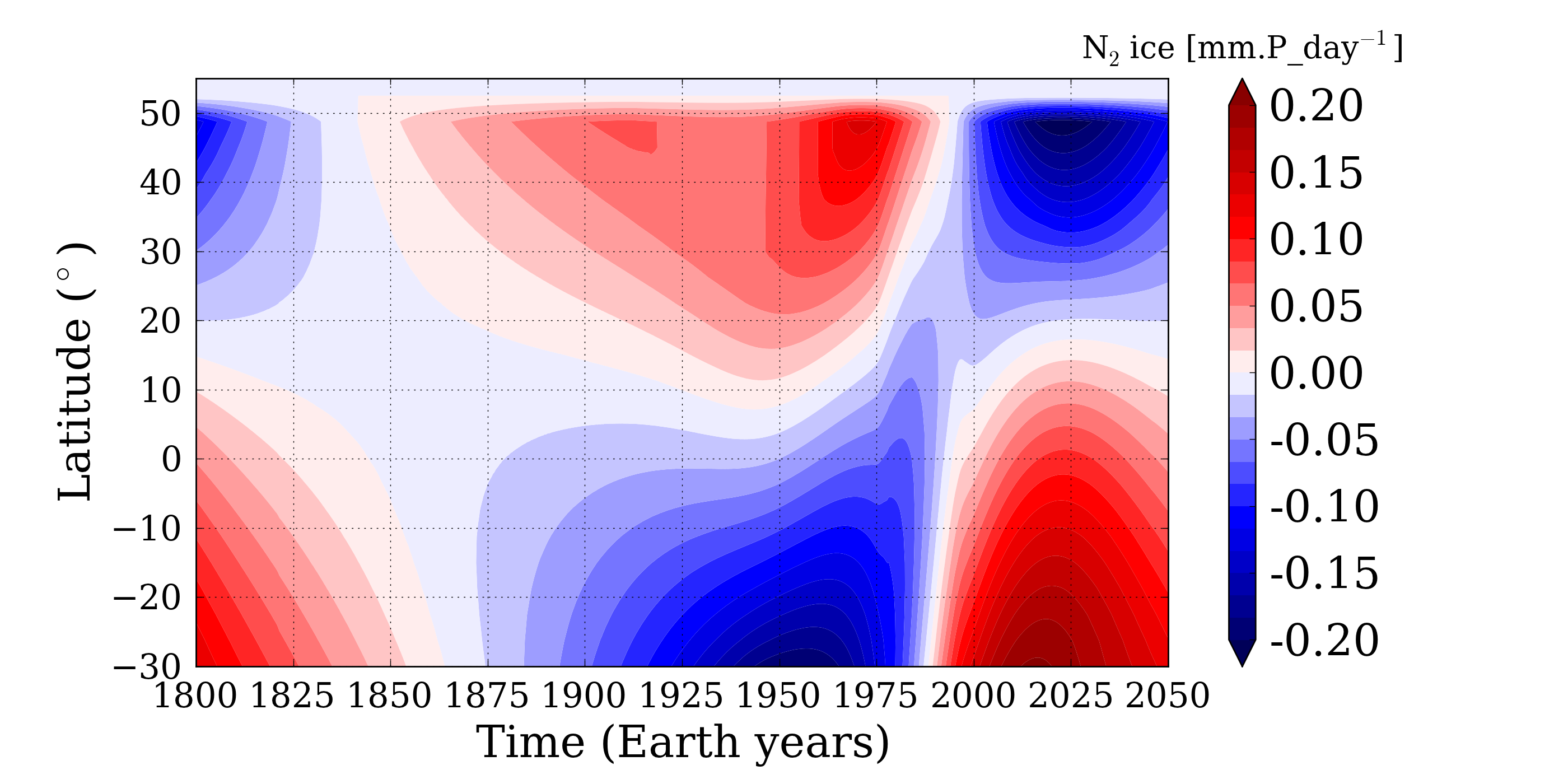}
\end{center} 
\caption{Evolution of the diurnal mean condensation-sublimation rate within Sputnik Planitia (mm per Pluto day), in current orbital conditions, from 1800 to 2050 assuming that the glacier remains flat.} 
\label{evol2dseason}
\end{figure}

\begin{figure}[!h]
\begin{center} 
	\includegraphics[width=16cm]{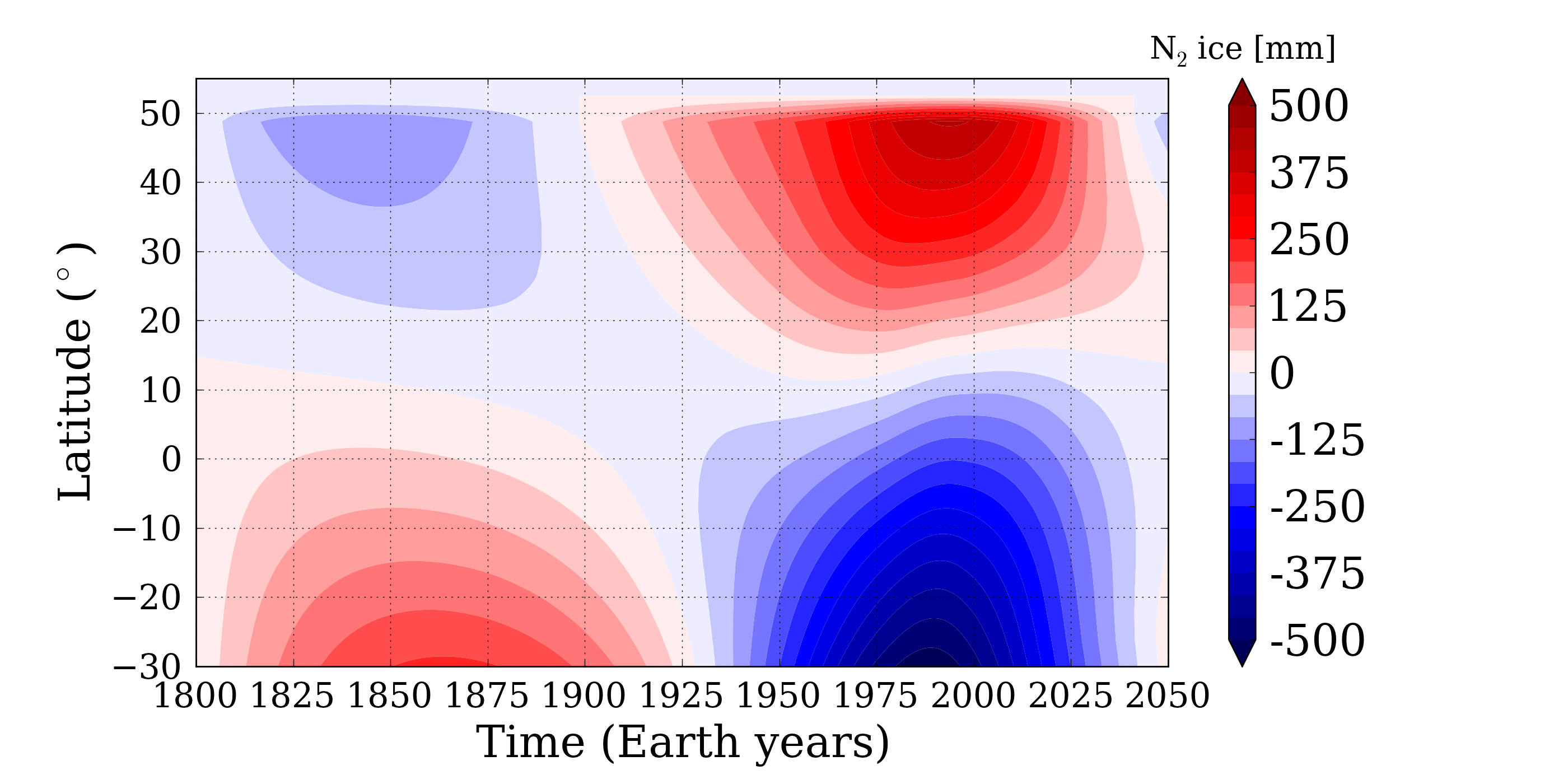}
\end{center} 
\caption{Variations of N$_2$ ice thickness within Sputnik Planitia (normalized to 0 at t=1800), in current orbital conditions, from 1800 to 2050 assuming that the glacier remains flat. The data is taken at the longitude 180$^{\circ}$. Although the net budget of ice within one Pluto year varies by tens of mm (\autoref{maps}.B), the thickness of ice involved during this year reaches hundreds of mm.} 
\label{evol2dseason_brut}
\end{figure}

\vspace{1cm}

\begin{figure}[!h]
\begin{center} 
	\includegraphics[width=16cm]{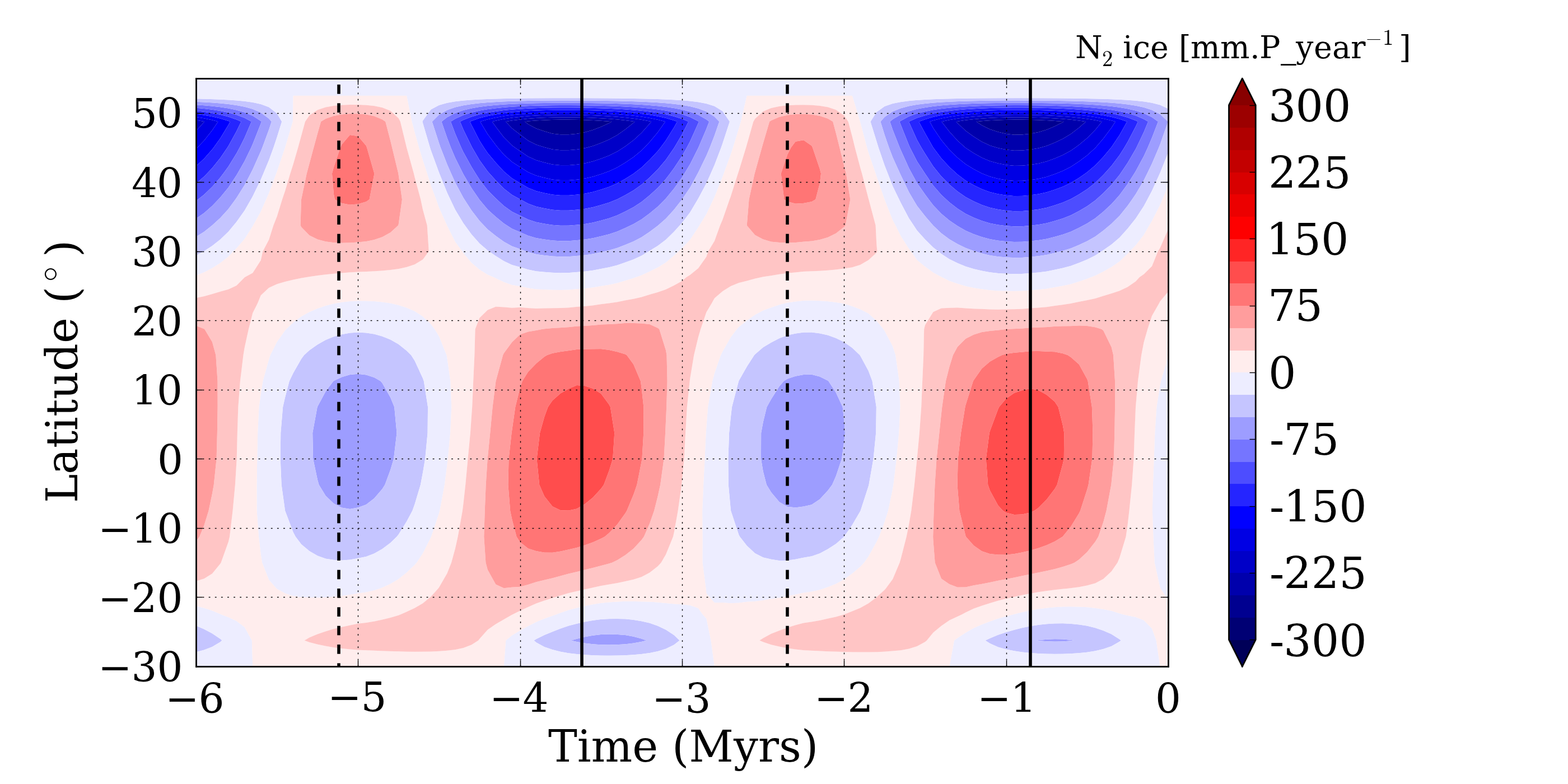}
\end{center} 
\caption{Evolution of the annual mean condensation-sublimation rate of N$_2$ ice with time (mm per Pluto year), assuming that the glacier remains flat (same as \autoref{evol2dseason}. The vertical solid and dashed lines correspond to the periods of high (104$^{\circ}$) and low (127$^{\circ}$) obliquity respectively.} 
\label{evol2d}
\end{figure} 

\clearpage

\subsubsection{The current annual timescale}

\autoref{evoldiurn_brut} shows the normalized diurnal variations of N$_2$ ice thickness over a Pluto day in July 2015. 
\autoref{evol2dseason} shows the evolution of the diurnal mean condensation-sublimation rate (net change of N$_2$ ice thickness) within Sputnik Planitia, over one current Pluto year, since 1800, while \autoref{evol2dseason_brut} shows the normalized variations of N$_2$ ice thickness at same dates (gross change of N$_2$ ice thickness). 

Over one Pluto day, several tens of micrometres of nitrogen ice move from the summer (North) to the winter (South) parts of SP (\autoref{maps}.A, \autoref{evoldiurn_brut}), while in one current Pluto year, a net amount of 20-50~mm of ice accumulates around 30$^{\circ}$N, 10-15~mm are lost in the southern part ($<$ 10$^\circ$N) and 20-50 mm are lost in the northern edge of SP ($>$ 40$^\circ$N, \autoref{maps}.B).

In 2015, the regions above 15$^{\circ}$N are in a sublimation-dominated regime, while regions below 15$^{\circ}$N are in a condensation-dominated regime (\autoref{maps}.A, \autoref{evol2dseason}). The southern regions of SP entered the condensation-dominated regime after the northern spring equinox in 1988, where a fast transition of regime between the northern and southern regions occurred. Before 1988, the southern regions had been in a sublimation-dominated regime since 1865. 
The net variation of ice thickness after one Pluto year reaches tens of mm (\autoref{maps}.B, \autoref{evol2dseason}) but the sublimation-condensation during this period involves thicker layers of ice (by a factor 10-30, \autoref{evol2dseason_brut}). Between 1865 and 1988 (7033 Pluto days), the southern regions lost 0.3-1~m of N$_2$ ice. 
Between 1988 and 2015, the regions below 10$^\circ$S accumulated 0.15-0.25~m of N$_2$ ice. 

\textbf{Association with the observed pits south of Sputnik Planitia}
\autoref{evol2d} shows the annual mean condensation-sublimation rates over the last 6 Myrs. 
The southern latitudes of SP (20$^{\circ}$S-10$^{\circ}$N) are in a sublimation-dominated regime since at least 100~000 years. The region below 20$^{\circ}$S is a sublimation-dominated regime since 1.3 Myrs and currently starts to enter a condensation-dominated regime.
The net loss of ice involved at the annual timescale in these regions occurs in the model at the same latitudes where the small pits are observed, explaining their formation there if they formed by sublimation.
Between 20$^{\circ}$S-10$^{\circ}$N, if we assume relatively similar insolation conditions over the last 100~000 Earth years, with a mean net sublimation rate of 15~mm of N$_2$ ice per Pluto year (\autoref{maps}.B), then the total loss of ice in this region could reach 6~m. Below 20$^{\circ}$S, assuming a mean sublimation rate of 50~mm of N$_2$ ice per Pluto year (\autoref{evol2d}) over 1.3 Myrs, the loss of ice reaches $\sim$~260~m. These values are in accordance with the observed depth of the pits (tens of meters, \citet{Moor:17}). Other mechanisms not taken into account in this model, such as atmospheric winds, light reflection and deposition of dark materials at the bottom of the pits may further increase the sublimation rate and favor deeper pits formation. 
This annual mean sublimation pattern could also explain the disappearance of the polygonal cells (if they are erased by sublimation), although this may also be related to a lower ice thickness (too low to trigger convection), as it is probably the case for all edges of SP. 

\begin{figure}[!h]
\begin{center} 
	\includegraphics[width=12cm]{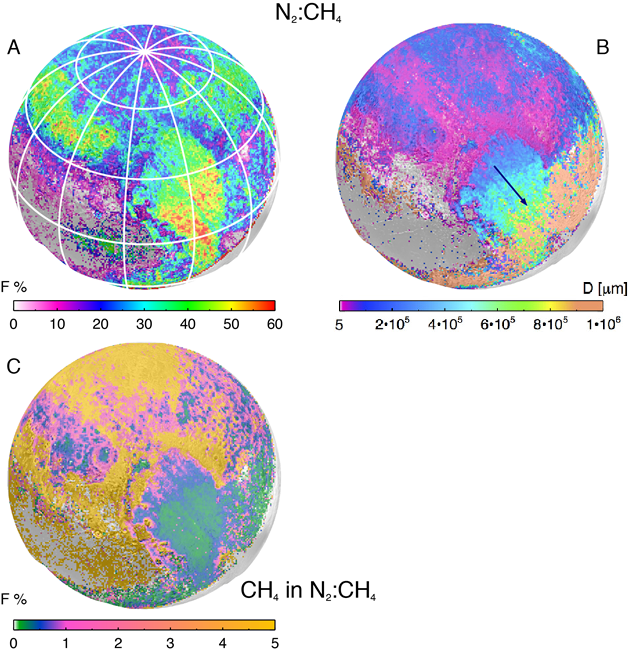}
\end{center} 
\caption{Modeling results from \citet{Prot:17} showing the abundance (A) and the path length (B) of the N$_2$-enriched. Panel C shows the dilution content of CH$_4$ in N$_2$. } 
\label{silvia}
\end{figure}

\textbf{Association with the ice albedo and composition}
Our results also show that the latitudes where N$_2$ ice accumulates in average over one Pluto year (\autoref{maps}.B) correspond to the latitudes where bright N$_2$ plains and a weaker amount of CH$_4$ in the N$_2$-CH$_4$ mixture (both are correlated) are observed in SP \citep{Prot:17,Schm:17,Bura:17}. 
\citet{Prot:17} note that the abundance of N$_2$:CH$_4$ is higher at the center of Sputnik Planitia with respect to the northern area of the basin, contrary to the dilution content of CH$_4$ in the mixture (\autoref{silvia}). They interpret this trend in the composition maps as a possible north-south sublimation transport of nitrogen in Sputnik Planitia (indicated schematically by the arrow in panel B of \autoref{silvia}). This is now supported by our results showing a net deposition of N$_2$ ice in the middle of the basin over the seasonal timescales (\autoref{maps}.B), and a recent deposition of few micrometers of N$_2$ ice in the southern latitudes during the past 30 Earth years (\autoref{evol2dseason}).

\vspace{1cm}

Despite its net daily sublimation since about 30 Earth years (\autoref{evol2dseason}), the darker cellular plains are currently an area of net ice accumulation on the annual timescale (about 4-8 m in the last 100~000 Earth years, \autoref{maps}.B and \autoref{evol2d}), explaining why they also remain relatively bright compared to the northern dark trough-bounding plains (above 40$^{\circ}$N), which are subjected to net annual sublimation since almost 1.8 Myrs (\autoref{evol2d}).

\vspace{0.5cm}
\subsubsection{The astronomical timescale}

\autoref{maps}.D shows that over the last obliquity cycle (2.8 Myrs ago up to now), up to 300 m of N$_2$ ice accumulated between 20$^{\circ}$S-30$^{\circ}$N, while an intense loss of about 1 km of ice occurred at the northern edge of the ice sheet between 30$^{\circ}$N-50$^{\circ}$N. In addition, at the southern edge of SP (below 20$^{\circ}$S), a net loss of 150 m of ice also occurred.
As shown by \autoref{evol2d}, N$_2$ sublimation at the northern edge of SP is the most intense during the periods of high obliquity (e.g 0.85 Myrs ago), and still occurs there during a large part of the obliquity cycle, for obliquities lower than 119$^{\circ}$ (higher than 61$^{\circ}$ in retrograde rotation).
As shown by \autoref{maps}.C, during the last 0.5 Myrs, the mean accumulation and loss of ice occurred at similar latitudes than during the last 2.8 Myrs, except between 15$^{\circ}$S-20$^{\circ}$S since these latitudes are currently transitioning to a sublimation-dominated regime. During the last 0.5 Myrs, the center of SP accumulated up to 100~m of ice while the northern regions lost 200-300~m of ice, that is one third of what they have lost in average over the last obliquity cycle (2.8 Myrs). Note that a net loss of ice continuously occurs at the northern edge of SP since the last 1.8 Myrs. During the same period, ice has been continuously condensing between 20$^{\circ}$S-25$^{\circ}$N. 
We can associate several structures of SP to the change of N$_2$ ice thickness averaged over this astronomical timescale. 

\textbf{Depressions and outward glacial flows at the northern and southern edge of Sputnik Planitia}
First, the latitudes where 1-2 km deep depressions are observed at the northern and southern boundaries of the ice sheet (see Figure 8 and 17 in \citet{Howa:17}) coincide with the latitudes where intense sublimation of ice occurred in the last 2.8 Myrs. This loss of ice should tend to be compensated by glacial flow, in line with the outward direction of the flow observed at these edges, and with the evidences of the particularly strong erosion of the Al-Idrisi Montes at the northern edge of SP \citep{Howa:17}. Simulations with glacial flow are explored in Section~\ref{secpaleo:cyclesflow}.

\textbf{Recent glacial activity at the eastern side of Sputnik Planitia}
Secondly, the recent glacial activity of ice flowing westward through the valleys of the eastern side of SP (pink color on \autoref{maps}.F, the observations show that the ice at the eastern side of SP flows toward the center of SP) occurs at the same latitudes where nitrogen ice continuously accumulated during the last 1.8 Myrs (20$^\circ$S-30$^\circ$N). 
We suggest that the accumulation of ice at these latitudes created a topography gradient at the edge of SP as the thick layer of ice far from the edge (closer to the center of SP) flowed more rapidly than the shallow one at the edge. These glacial flows should be reduced or disappear during the next hundreds of thousand years since these latitudes gradually enter a sublimation-dominated regime (\autoref{evol2d}). 

Why are such glacial flows not observed on the western side of SP, where the latitudes receive the same insolation? 
Glacial activity on the western side of SP has occurred as evidenced by the many erosional valleys, but these valleys are not filled with flowing ice like they are to the east. This could be due to the significant difference of geology between the western and eastern side. As shown by \citet{Sche:18}, a North-South giant fault system passes under the western edge of the ice sheet, which may explain the fragmentation of water ice blocks and the presence of deep ridge, faults, and cliffs observed on the western edge of SP. This topography may prevent the ice from flowing easily through the west side of SP and form large glacial flows. In addition, the western side of SP may correspond to a deeper bedrock than on the  eastern side, preventing strong gradients of nitrogen ice thickness and the inward flow observed in the eastern raising valleys (the ice would locally flow faster on the western edge of SP). 



\textbf{The dark northern plains of Sputnik Planitia}
The dark and methane rich aspect of the northern edge of Sputnik Planitia (40$^{\circ}$N-50$^{\circ}$N) is also consistent with our results, which show that this area is a sublimation area at all timescales: the current diurnal, the current annual, the last 0.5 Myrs and the last 2.8 Myrs. Why does this area not have small pits, like in the southern part of SP? A suggestion is that the area lacks an intake of fresh nitrogen ice necessary for the formation of pits. This may be because the area is subject to a net loss of ice at all timescales and the layer of ice became too shallow to undergo solid-state convection. In addition or alternatively, the methane rich composition and the size of grains of this dark ice may play a restricting role in the formation of pitted plains. 
Finally, the formation of pits is a process of erosion by reflected light. If the ice albedo is too low, the direct absorption of solar energy predominates the reflection and the ice sublimates uniformly, inhibiting the pit formation \citep{Moor:17}.   

\subsection{The astronomical cycles with glacial flow}
\label{secpaleo:cyclesflow}

In this section, we repeat the same simulations as in Section~\ref{secpaleo:cycles} except that we turn on the glacial flow scheme of N$_2$ ice (described in Section~\ref{secpaleo:flow}), which enables the ice to flow in the modelled Sputnik Planitia basin (Section~\ref{secpaleo:topo}). The basin is initially filled with N$_2$ ice up to 2.5~km below the mean surface level. We assume that the edges of the basin are at about 3~km, 4~km or 5~km below the mean surface level (\autoref{topopluton}).

\subsubsection{General overview}
As a general tendency, our results show that over one obliquity cycle, only small variations of elevation up to 25~m are obtained in the centre of SP at 20$^{\circ}$N (SP remains relatively flat where the ice is thick) while variations of elevations of 200-300 m are obtained at the edges of SP (\autoref{evolflow1d} and \autoref{mapflow2d}), in particular at the northern and southern edges. The variations are reduced if we assume the bedrock deeper below the ice sheet.
These variations of elevation obtained with our model are consistent with the depressions observed at the northern and southern boundaries of SP and with the eroded mountain blocks observed west and east of Tenzing Montes, indicative of the presence of large amounts of ice there in the past.
\autoref{evolflow1d} suggests that the ice sheet was at its maximal North-South extension 1.5-2 Myrs ago, since the ice level in the Al-Idrisi region and in the far south of SP was well above the level of the centre of SP. Conversely, the last million of years coincides with a period of minimal extension, which is consistent with the ice flowing outward from SP at its northern and southern edges \citep{Howa:17} and the lower MVIC-derived topography observed north of the Al-Idrisi region \citep{Sche:17AGU}.

\subsubsection{Comparisons with observations}

\begin{figure}[!h]
\begin{center} 
	\includegraphics[width=15cm]{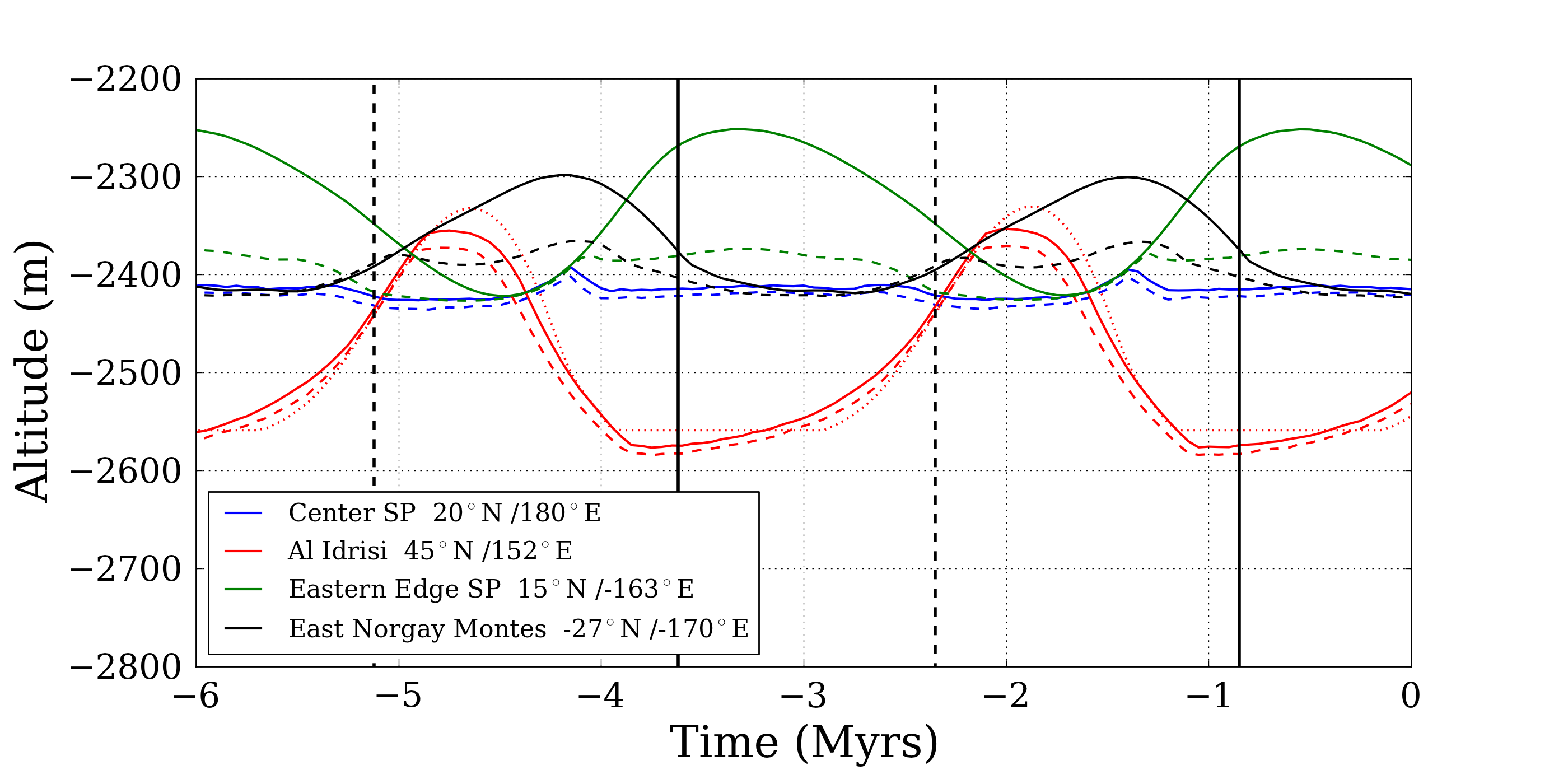}
\end{center} 
\caption{Variations in elevation within SP at different locations, assuming that the bedrock below SP include a 6-9 km deep elliptical basin and  edges at 4~km (solid lines) and 5 km (dashed lines) below the mean surface level. The red dotted line corresponds to a case with edges at about 3 km below mean level. In that case, the ice has been entirely sublimed during the last millions of years at Al-Idrisi, as the elevation shown during that time is the bedrock level at this location (flat line at 2.55 km below the mean surface). The vertical solid and dashed lines correspond to the periods of high (104$^{\circ}$) and low (127$^{\circ}$) obliquity respectively. \newline} 
\label{evolflow1d}
\end{figure}

\begin{figure}[!h]
\begin{center} 
	\includegraphics[width=8cm]{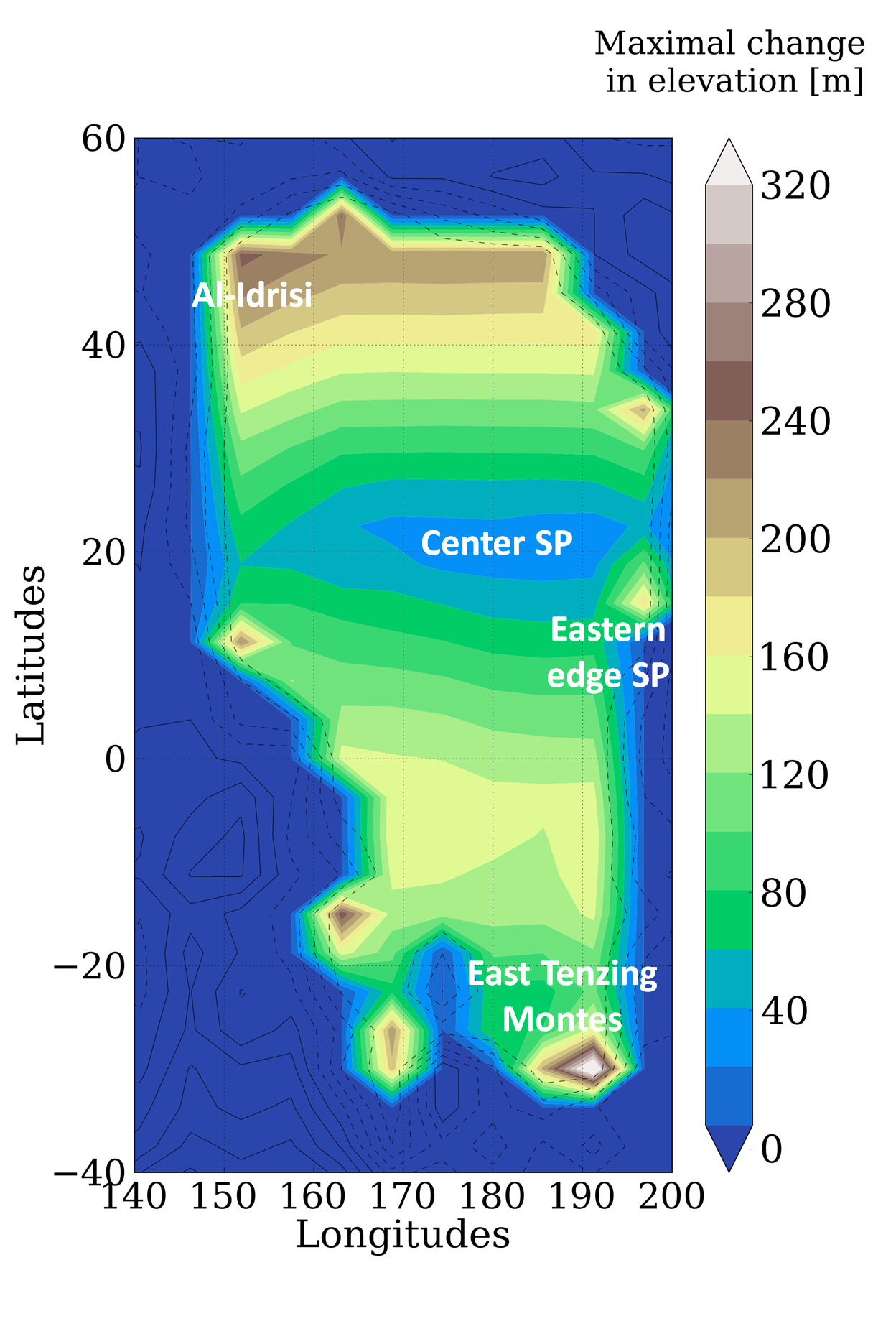}
\end{center} 
\caption{Maximal variation in elevation of N$_2$ ice over the last obliquity cycle (last 2.8 Myrs), with a bedrock on the edges of SP at 4~km and an initial filling of SP at 2.5~km. \newline} 
\label{mapflow2d}
\end{figure}

\textbf{The Al-Idrisi Montes}
As shown by \autoref{mapflow2d}, the entire northern part of SP (above 40$^{\circ}$N) is subject to variations of altitudes of 100-280~m, which is consistent with the intense sublimation and condensation of N$_2$ ice occurring over one obliquity cycle at these latitudes (\autoref{maps}.C-D). 
In particular, the latitudes of the Al-Idrisi Montes displays one of the largest variations of elevations (up to 280~m), in agreement with the scenario of strong and endless erosion of the water ice blocks in this region (\autoref{evolflow1d} and \autoref{mapflow2d}). In the simulation with the less deep bedrock on the edges of SP, the ice at the latitudes of the Al-Idrisi Montes sublimed during the last 2 Myrs and revealed the bedrock (which is 2.55 km below the mean surface at this location in that case, \autoref{evolflow1d} red dotted line). Currently, this region enters a regime of net accumulation over astronomical timescales. 

\textbf{The recent glacial activity at the eastern side of Sputnik Planitia}
\autoref{evolflow1d} also shows the evolution of the ice at the eastern edge of SP, at 15$^{\circ}$N. As predicted by \autoref{evol2d}, the area accumulated up to 150~m of ice during the high obliquity periods, if the bedrock is at 4~km below mean surface (the variation is less if the bedrock is deeper). The elevation of this area decreases since 0.6~Myrs because the flow of ice toward the center of SP overcomes the intake of nitrogen from condensation. Note that this area is always higher than the center of SP. Thus, the glacial flows induced from the uplands to the center of SP, as observed by New Horizons, should never stop.  

\textbf{The southern latitudes of Sputnik Planitia}
The south of SP also displays strong variations of elevation, that are about 200~m over one obliquity cycle, with a bedrock at 4~km below mean surface (\autoref{mapflow2d}). At the east of the Tenzing Montes (27$^{\circ}$S,-170$^{\circ}$E), the elevation of the ice is higher than the center of SP during most of the obliquity cycle. 

\section{Possible steady-state conditions for Pluto's ices}
\label{secpaleo:eqstates}

In this section, we explored the stability of N$_2$ ice deposits outside Sputnik Planitia. To do that, we performed several simulations using different sensitivity parameters and initial states. 

\subsection{Simulation settings}

We used the following settings for the simulations: 
(1) We performed the simulations over the last 30 Myrs, taking into account the obliquity and orbital changes over time described in Section~\ref{orbitchanges}.
(2) We used realistic reservoirs of N$_2$ ice corresponding to a global surface coverage of Pluto of 200~m, 500~m or 1000~m of ice. The case of 500~m corresponds to a basin of 1200x1000~km filled by $\sim$~7~km of ice, which is in the range of what is assumed for SP.
(3) At the beginning of these simulations 30 Myrs ago, the surface is not initialized with the entire N$_2$ reservoir trapped inside SP as in Section~\ref{secpaleo:SP}. Instead, the initial N$_2$ reservoir is either globally uniformly distributed (Simulations $\sharp$Glob), or placed at the equatorial regions between $\pm$~30$^{\circ}$ latitude (Simulations $\sharp$Equa), or at the poles above 50$^{\circ}$ latitude (Simulations $\sharp$Polar). As an example, an initial global reservoir of 500~m redistributed over the equatorial regions between $\pm$~30$^{\circ}$ latitude corresponds to an initial equatorial reservoir of $\sim$1~km of ice. 
(4) We used the latest topography data from New Horizons coupled with a deep bedrock for SP (up to 10~km deep), as described in Section~\ref{secpaleo:topo}, and the glacial flow scheme described in Section~\ref{secpaleo:flow}. 


The sensitivity parameters of the simulations are the following: 
(1) Seasonal thermal inertia of 400, 800 and 1200 SI are used, similar than those used in \citet{BertForg:16}. The diurnal thermal inertia remains fixed at 20~SI \citep{Lell:11b}, as in \citet{BertForg:16}. 
(2) The reference N$_2$ albedo and emissivity used are set to 0.7 and 0.8 respectively, while those for bare ground are set to 0.1 and 1 respectively, which is in the range of what has been used in \citet{BertForg:16}, assuming the water ice bedrock is covered by dark tholins. We also explored the case of an albedo of 0.4 for N$_2$ ice. 

\subsection{Simulation results}
\label{secpaleo:results}

The results are summarized in Table~\ref{tab:results} and illustrated by \autoref{uniform1} and \autoref{uniform2}.

\vspace{0.5cm}
\subsubsection{Overall outcome}

As a general rule, N$_2$ ice quickly accumulates in the Sputnik Planitia basin and in the equatorial regions (preferentially at latitudes around $\pm$ 30$^{\circ}$), with stronger condensation rates inside SP and inside other depressions because of the stronger infrared cooling effect, as detailed in Section~\ref{secpaleo:intro} and \citet{BertForg:16}. 

However, in many of the simulations, large deposits also remain in the equatorial regions outside SP after 30 Myrs, and even beyond as they seem to remain relatively stable with time. 
The simulation $\sharp$Polar8 described in Table~\ref{tab:results} and shown on \autoref{uniform1} provides a typical example. Starting with an initial global reservoir of 500~m  confined at the poles and a thermal inertia of 1200~SI, the ice migrates toward the equatorial regions by forming latitudinal bands which get closer to the equator with time. 
The basin SP is progressively filled by N$_2$ ice, with a decreasing rate with time, because the ice outside SP migrates towards the more stable equatorial regions, leading to lower condensation rates inside SP, and also because as the surface of the basin becomes less deep, the infrared cooling effect becomes less efficient. 
The parameter $\tau^{SP}_{95\%}$ indicates the time needed to fill SP at 95$\%$ of its final state (in Myrs). It depends on the TI, the reservoir and the initial state. The lower this time, the more stable are the deposits outside SP. As an example, in the simulation $\sharp$Polar8, the basin is already in a relatively stable state after 11.90 Myrs. After 30 Myrs, it is filled by ice up to 2350 m below the mean surface (Table~\ref{tab:results}). Outside SP, N$_2$ ice remained at the equator forming 600-800~m deposits. After these 30 Myrs, the ice still migrates in the basin because of the infrared cooling effect but at a very slow rate. Typically, in 1 Myrs, the equatorial deposits lost 5-10 meters of ice. Consequently, the 600-800~m deposits outside SP should end trapped inside SP after at least 60 additional Myrs.

We note that the ice in the equatorial regions outside SP is sligthly less stable for L$_{s~peri}$ values close to 90$^{\circ}$ and 270$^{\circ}$ (values favouring an asymmetry of surface temperatures between both hemisphere). 
Finally, N$_2$ ice is never stable at the poles and any initial polar deposit up to 1~km thick is entirely sublimed after less than 2 Myrs. 

\vspace{0.5cm}
\subsubsection{Sensitivity to the ice reservoir}

The larger the ice reservoir, the faster the glacial flow and the more easily the ice reaches the equatorial regions (outside SP), where it is able to form relatively stable deposits several hundred meters deep. 
In addition, larger reservoirs of ice lead to larger N$_2$ ice deposits outside SP, spread from the equator toward higher latitudes.

In our simulations, equatorial deposits outside SP inevitably form as soon as the initial reservoir is equal to or larger than 500~m, independently of any initial distribution as long as it is assumed that the initial reservoir is outside SP.
If the initial reservoir is lower, the ice does not flow easily and the presence of equatorial deposits outside SP depends on the TI and the initial state (see section \ref{secpaleo:sensibTI}).

As an example, in the simulations $\sharp$Polar4 and $\sharp$Polar7 performed with a reservoir of 200~m of ice, all the ice ends in the SP basin after 30 Myrs (see \autoref{uniform1}) and fills it up to 3.29 km below the mean level. 
If the 200~m of ice are initially present at the equator (Simulation $\sharp$Equa and $\sharp$Glob), and if the TI is higher than 800~SI then 200-400~m thick ice deposits can persist in the equatorial regions after 30 Myrs (e.g. simulations $\sharp$Glob4, $\sharp$Glob7, $\sharp$Equa4, $\sharp$Equa7, see Table~\ref{tab:results}). 

While in the simulations performed using a reservoir of 200~m the SP basin is filled by ice up to 3.29 km below the mean level after 30 Myrs, in all simulations using a reservoir of 500~m of ice, the SP basin is filled up to 1500-2300~m below the mean level.
In all simulations using a reservoir of 1~km of ice, the basin is entirely filled with ice. In these cases, the basin fills up very rapidly because a large amount of ice is able to flow directly inside the basin. 

\vspace{0.5cm}
\subsubsection{Sensitivity to thermal inertia and initial state}
\label{secpaleo:sensibTI}

Our results are sensitive to the assumed thermal inertia. As shown by \autoref{fluxtemp} and \autoref{tempmean}, the lower the TI, the less equatorial are the cold points on Pluto's surface (in average over the last Myrs). In particular, if TI is lower than 800 SI, the equator becomes warmer in average than the other latitudes. In the simulations using a reservoir of 200~m of ice, different final states are obtained after 30 Myrs depending on the TI and the initial state.


(1) If the initial reservoir is distributed at the poles ($\sharp$Polar1, $\sharp$Polar4, $\sharp$Polar7), we found that the ice subsists outside SP after 30 Myrs only if the TI is equal to or lower than 400~SI, where a relatively stable latitudinal band of 330~m N$_2$ ice can form at 20$^{\circ}$N ($\sharp$Polar1, \autoref{uniform1}). 
If the TI is larger ($\sharp$Polar4, $\sharp$Polar7), the entire reservoir is quickly trapped in SP (\autoref{uniform1}).  

(2) If some of the initial reservoir is distributed at the equator ($\sharp$Equa, $\sharp$Glob), the ice can remain outside SP after 30 Myrs only if the TI is equal to or larger than 800 SI. The ice forms latitudinal bands of 300-800~m N$_2$ ice at the latitudes $\pm$~10$^{\circ}$, with higher amounts in the local depressions (see \autoref{uniform2}, $\sharp$Glob4 and $\sharp$Glob7).
If the TI is lower ($\sharp$Glob1, $\sharp$Equa1), the ice is less stable at the equator and ends trapped in the SP basin (\autoref{uniform2}).
Generally speaking, results obtained using an initial ice reservoir distributed at the equator and over the globe are similar and therefore we only show the latter results on \autoref{uniform2}.
Note that when larger TI are used, larger amounts of ice remain in the equatorial regions. 
In addition, the deposits tend to peak at the equator for large TI, while they tend to peak at higher latitudes ($\pm$10$^{\circ}$, or $\pm$20$^{\circ}$) for low TI (see e.g. $\sharp$Polar1, $\sharp$Glob4).


\vspace{0.5cm}


\vspace{0.5cm}
\subsubsection{Sensitivity to the albedo}

The simulations using a reservoir of 500~m of ice have also been performed using an albedo of 0.4 for N$_2$ ice ($\sharp$Polar10-12, $\sharp$Equa10-12, $\sharp$Glob10-12), instead of 0.7 (reference value). In all these low albedo simulations, the ice sublimates very rapidly from the poles and accumulates in the equatorial regions between 37.5$^{\circ}$S and 37.5$^{\circ}$N (the ice is sligthly more spread than in the cases with an albedo of 0.7), forming stable deposits about 600-800~m deep. These results are found to be relatively independent of the initial state and of the thermal inertia. 
This is because the lower albedo enables the ice to be warmer and to gain greater mobility (by both glacial flow and condensation-sublimation flux) to reach the coldest point on Pluto's surface. 

In these simulations, SP is rapidly filled by ice and reaches a relatively stable level. After $\tau^{SP}_{95\%}$=7-9 Myrs, the elevation of SP surface then increases by only a few meter every Myrs due to N$_2$ ice condensation. 

\vspace{0.5cm}
\subsection{Minimum and maximum surface pressures}
\label{secpaleo:pressures}

\autoref{evolpres} shows the evolution of the maximum and minimum annual surface pressures during the last 15 Myrs obtained in  simulations $\sharp$Polar1, $\sharp$Polar4, $\sharp$Polar8 and $\sharp$Polar12.
Generally speaking, the surface pressures (and surface temperatures) remain within 10$^{-2}$-10~Pa (31-40~K) in all simulations using an albedo for N$_2$ ice of 0.7, and within the range 1-100~Pa (39-45~K) in all simulations using an albedo for N$_2$ ice of 0.4 (in the model we are always in the global-atmosphere regime).
Higher maximum surface pressures could be obtained, if we lower the albedo below 0.4 or the TI below 400~SI, but such values seem quite distant from reality.   
We note that (1) there are two peaks of maximum surface pressure per obliquity cycle (2) Maximum pressures are lower during the high obliquity periods (104$^{\circ}$). This is because the main reservoirs of ice are located in the equatorial regions, which receive less flux on annual average during these periods (\autoref{fluxtemp}) (3) The surface pressures in the simulations with N$_2$ ice deposits outside SP are sligthly less than in the simulation without deposits outside SP. 

\begin{figure}[!h]
\begin{center} 
	\includegraphics[width=15cm]{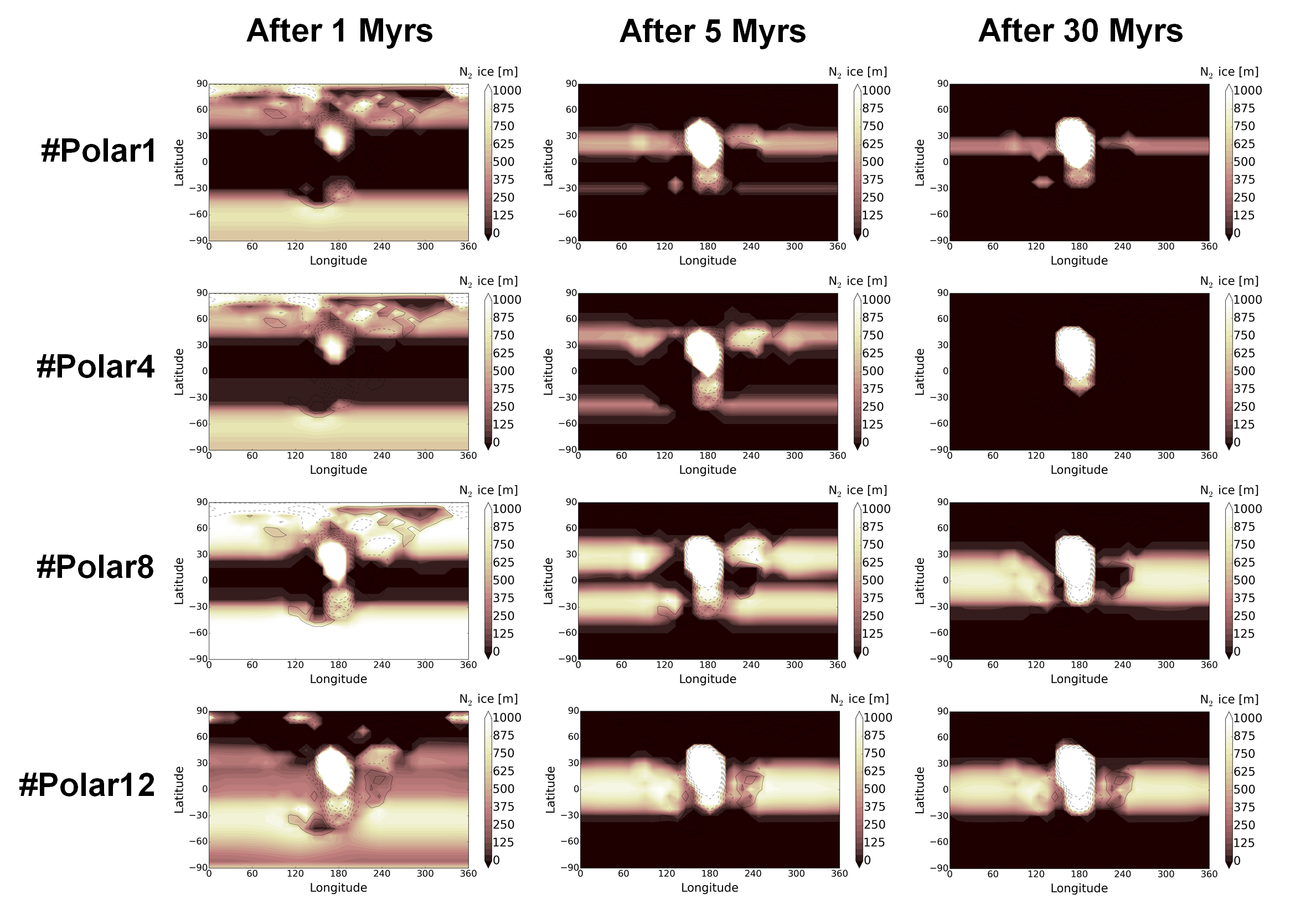}
\end{center} 
\caption{Maps of N$_2$ ice distribution on Pluto (m) for simulations starting 30~Myrs ago with a polar reservoir ($\sharp$Polar1, $\sharp$Polar4, $\sharp$Polar8, $\sharp$Polar12). Results are shown after 1~Myrs (left panel), 5~Myrs (middle panel) and 30~Myrs (right panel).\newline} 
\label{uniform1}
\end{figure} 

\begin{figure}[!h]
\begin{center} 
	\includegraphics[width=15cm]{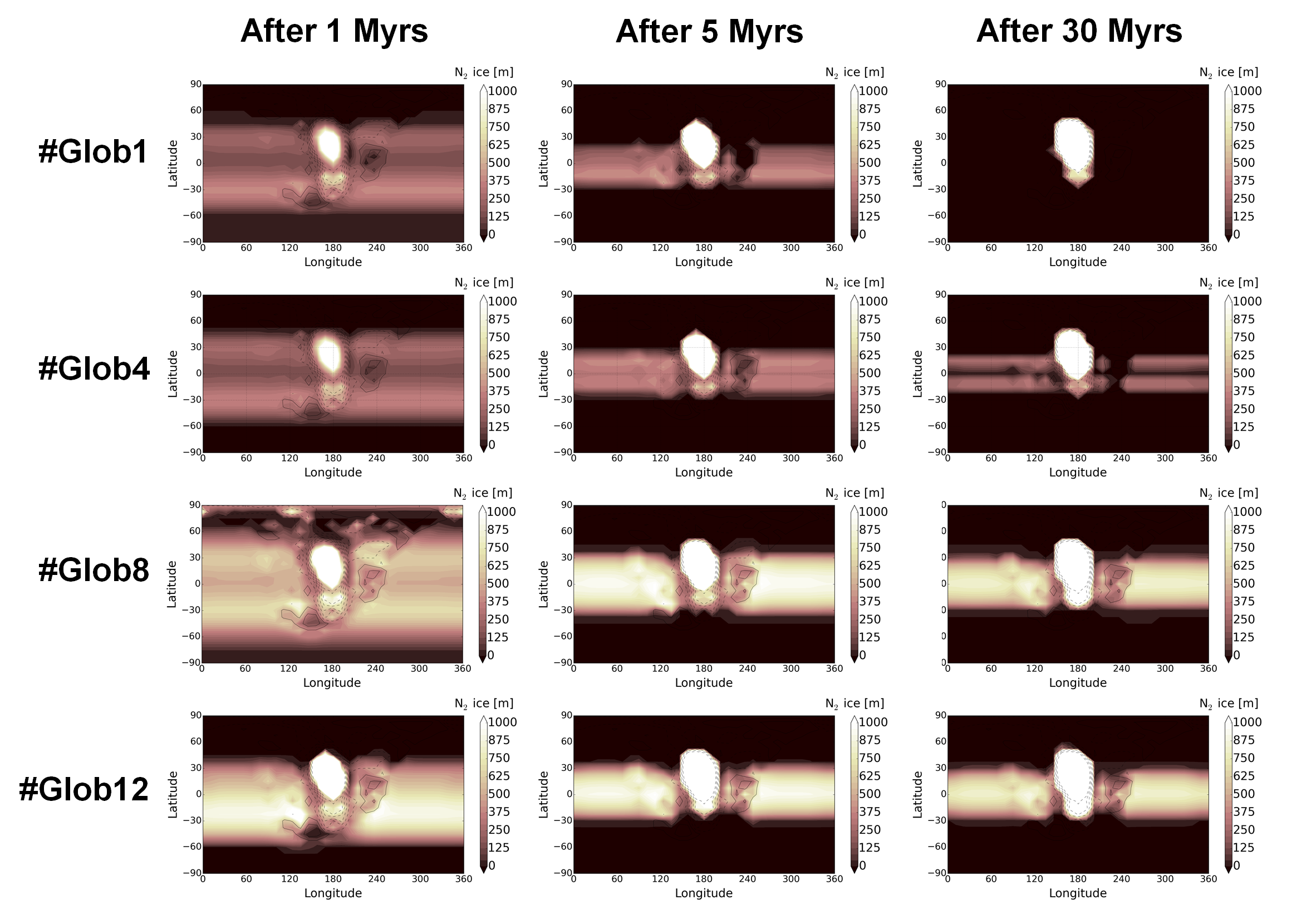}
\end{center} 
\caption{Maps of N$_2$ ice distribution on Pluto (m) for simulations starting 30~Myrs ago with a global reservoir ($\sharp$Glob1, $\sharp$Glob4, $\sharp$Glob8, $\sharp$Glob12). Results are shown after 1~Myrs (left panel), 5~Myrs (middle panel) and 30~Myrs (right panel).\newline} 
\label{uniform2}
\end{figure} 

\begin{figure}[!h]
\begin{center} 
	\includegraphics[width=15cm]{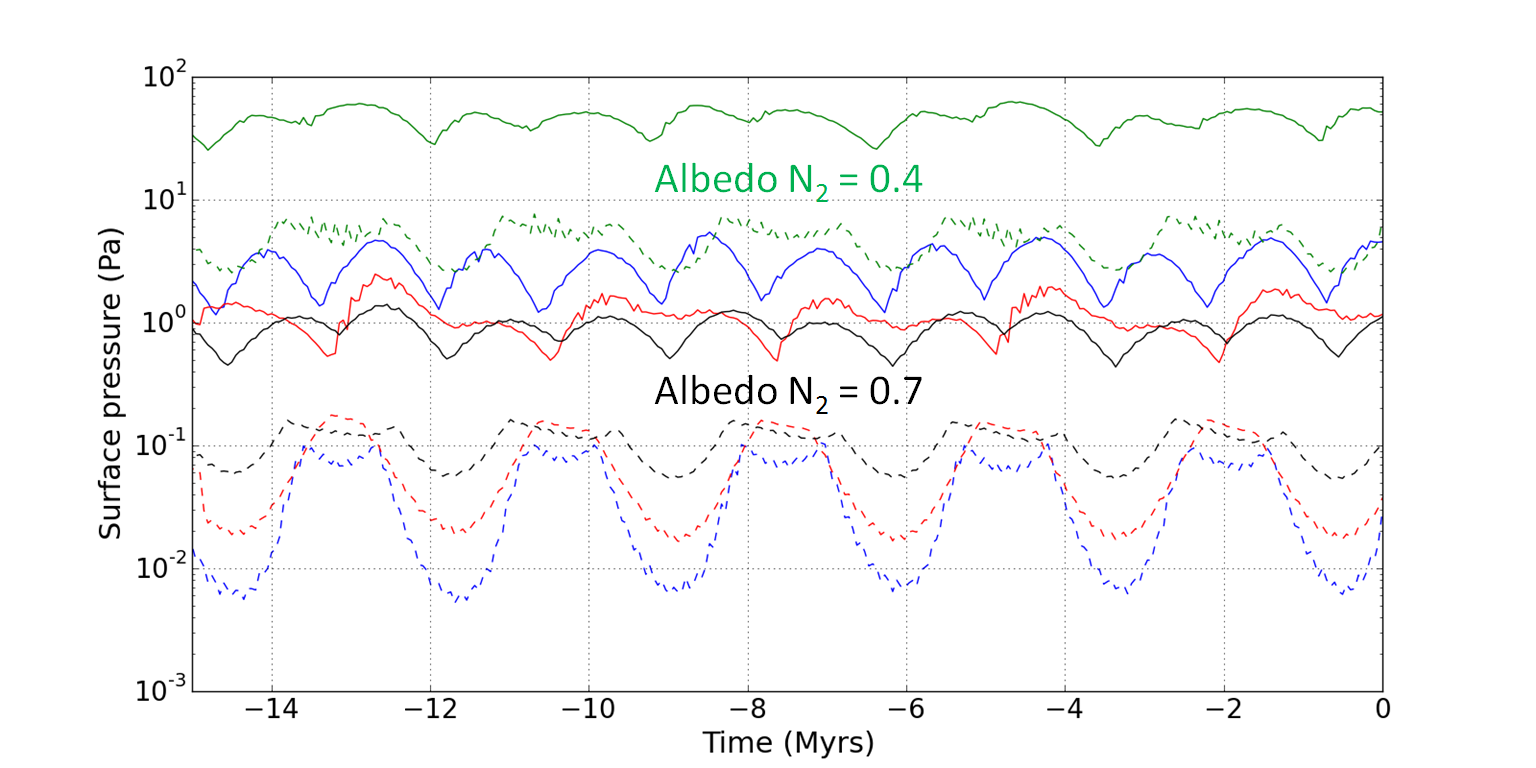}
\end{center} 
\caption{Evolution of maximum (solid lines) and minimum (dashed lines) annual surface pressure over the last 15~Myrs for simulations starting with a polar reservoir: $\sharp$Polar1 (blue), $\sharp$Polar4 (red), $\sharp$Polar8 (black), $\sharp$Polar12 (green). The present-day maximum surface pressure is $\sim$~1.1 Pa \citep{Ster:15,Glad:16,Hins:17}. \newline} 
\label{evolpres}
\label{lastfig}
\end{figure} 

\begin{sidewaystable}
\renewcommand{\arraystretch}{0.5} 
\centering
\caption{Settings and results of the simulations performed from 30 Myrs ago to present-day. From left to right, settings are: name of the run (the runs marked by * are illustrated by \autoref{uniform1} and \autoref{uniform2}), thermal inertia, N$_2$ ice reservoir (globally averaged), N$_2$ ice albedo. Results are: latitudes between which N$_2$ ice deposits are obtained outside SP, latitudes where the N$_2$ ice deposits are obtained outside SP peak, altitude of the N$_2$ ice deposits inside SP, altitude of N$_2$ ice outside SP, time needed to fill SP with N$_2$ ice at 95$\%$ of its final state, minimum and maximum surface pressures obtained during the last 15 Myrs}
\label{tab:results}
\label{lasttable}
\begin{tiny} 
\tabcolsep=0.11cm
\begin{tabular}{m{2cm}m{2cm}m{1.5cm}m{1cm}m{2cm}m{7cm}m{1.2cm}m{1.5cm}m{1cm}m{1cm}m{1cm}} 
\hline
\hline
   & & & & & & H$_{ice}$ & H$_{ice}$ & & \multicolumn{2}{c}{Surface Pressure} \\
   & Thermal inertia & Reservoir & Albedo & Latitude N$_2$ & Peak N$_2$ & in SP & outside SP & $\tau^{SP}_{95\%}$ & \multicolumn{2}{c}{(Pa)} \\ 
Run & (J~s$^{-1/2}$~m$^{-2}$~K$^{-1}$) &  (kg~m$^{-2}$) & A$_{N2}$ & outside SP & outside SP & (m) & (m) & (Myrs) &  P$_{min}$ & P$_{max}$ \tabularnewline 
\hline
\hline
\textbf{$\sharp$Polar1*} & 400 & 200 & 0.7 & 15$^{\circ}$N-22.5$^{\circ}$N & 18$^{\circ}$N & -3960 & 330 & 22.45 & 0.0056 & 4.945  \\
\textbf{$\sharp$Polar2} & 400 & 500 & 0.7 & 30$^{\circ}$S-30$^{\circ}$N  & 15$^{\circ}$S & -1540 & 400-600 & 10.15 & 0.0166 & 3.977 \\
\textbf{$\sharp$Polar3} & 400 & 1000 & 0.7 & 50$^{\circ}$S-50$^{\circ}$N  & 10$^{\circ}$S & 440 & 1000-1200 & 6.55 & 0.0234 & 2.371 \\
\textbf{$\sharp$Polar4*} & 800 & 200 & 0.7 & No ice & / & -3290 & 0 & 26.55 & 0.0229 & 2.355 \\
\textbf{$\sharp$Polar5} & 800 & 500 & 0.7 & 30$^{\circ}$S-30$^{\circ}$N & Equator & -2110 & 500-750 & 27.05 & 0.0371 & 2.101  \\
\textbf{$\sharp$Polar6} & 800 & 1000 & 0.7 & 50$^{\circ}$S-50$^{\circ}$N & Equator & 440 & 1000-1200 & 12.30 & 0.0523 & 1.527 \\
\textbf{$\sharp$Polar7} & 1200 & 200 & 0.7 & No ice & / & -3290 & 0 & 35.00 & 0.1498 & 2.862 \\
\textbf{$\sharp$Polar8*} & 1200 & 500 & 0.7 & 30$^{\circ}$S-30$^{\circ}$N & Equator & -2350 & 500-800 & 11.90 & 0.0513 & 1.393  \\
\textbf{$\sharp$Polar9} & 1200 & 1000 & 0.7 & 50$^{\circ}$S-50$^{\circ}$N & Equator & 430 & 1000-1200 & 12.50 & 0.0679 & 1.162 \\
\textbf{$\sharp$Polar10} & 400 & 500 & 0.4 & 30$^{\circ}$S-37.5$^{\circ}$N & 5$^{\circ}$N or locally in the depressions & -1740 & 600-800 & 7.45 & 0.8433 & 88.485 \\
\textbf{$\sharp$Polar11} & 800 & 500 & 0.4 & 30$^{\circ}$S-37.5$^{\circ}$N & Equator or locally in the depressions & -1820 & 600-800 & 7.50 & 1.8224 & 70.005 \\
\textbf{$\sharp$Polar12*} & 1200 & 500 & 0.4 & 30$^{\circ}$S-37.5$^{\circ}$N & Equator or locally in the depressions & -1850 & 700-800 & 7.55 & 2.5942 & 62.690 \\
\hline          
\textbf{$\sharp$Equa1} & 400 & 200 & 0.7 & No ice & / & -3290 & 0 & 19.15 & 0.0073 & 4.919 \\
\textbf{$\sharp$Equa2} & 400 & 500 & 0.7 & 30$^{\circ}$S-30$^{\circ}$N & 10$^{\circ}$S & -1460 & 500-700 & 9.75 & 0.0160 & 4.192  \\
\textbf{$\sharp$Equa3} & 400 & 1000 & 0.7 & 50$^{\circ}$S-50$^{\circ}$N & 18$^{\circ}$S & 450 & 1000-1200 & 7.40 & 0.0232 & 2.397 \\
\textbf{$\sharp$Equa4} & 800 & 200 & 0.7 & 20$^{\circ}$S-20$^{\circ}$N & 10$^{\circ}$S 10$^{\circ}$N, less ice at the equator & -5290 & 300-400 & 7.95 & 0.0241 & 2.218  \\
\textbf{$\sharp$Equa5} & 800 & 500 & 0.7 & 37.5$^{\circ}$S-37.5$^{\circ}$N & Equator & -2030 & 600-700 & 27.85 & 0.0320 & 1.975  \\
\textbf{$\sharp$Equa6} & 800 & 1000 & 0.7 & 50$^{\circ}$S-50$^{\circ}$N & Equator & 460 & 1000-1200 & 12.25 & 0.0527 & 1.536 \\
\textbf{$\sharp$Equa7} & 1200 & 200 & 0.7 & 20$^{\circ}$S-20$^{\circ}$N & 10$^{\circ}$S 10$^{\circ}$N, less ice at the equator & -5340 & 300-400 & 7.70 & 0.0375 & 1.381  \\
\textbf{$\sharp$Equa8} & 1200 & 500 & 0.7 & 30$^{\circ}$S-30$^{\circ}$N & Equator & -2160 & 700-800 & 30.35 & 0.0485 & 1.448  \\
\textbf{$\sharp$Equa9} & 1200 & 1000 & 0.7 & 50$^{\circ}$S-50$^{\circ}$N & Equator & 440 & 1000-1200 & 12.20 & 0.0704 & 1.184  \\
\textbf{$\sharp$Equa10} & 400 & 500 & 0.4 & 30$^{\circ}$S-37.5$^{\circ}$N & 5$^{\circ}$N or locally in the depressions & -1740 & 600-800 & 9.65 & 0.8430 & 88.483 \\
\textbf{$\sharp$Equa11} & 800 & 500 & 0.4 & 37.5$^{\circ}$S-45$^{\circ}$N & Equator or locally in the depressions & -1820 & 600-800 & 9.40 & 1.8220 & 70.006 \\
\textbf{$\sharp$Equa12} & 1200 & 500 & 0.4 & 30$^{\circ}$S-37.5$^{\circ}$N & Equator or locally in the depressions & -1850 & 700-800 & 8.10 & 2.5915 & 62.693 \\
\hline                   
\textbf{$\sharp$Glob1*} & 400 & 200 & 0.7 & No ice & / & -3290 & 0 & 17.15 & 0.0073 & 4.754 \\
\textbf{$\sharp$Glob2} & 400 & 500 & 0.7 & 30$^{\circ}$S-30$^{\circ}$N & 10$^{\circ}$S & -1500 & 500-650 & 11.45 & 0.0160 & 4.147  \\
\textbf{$\sharp$Glob3} & 400 & 1000 & 0.7 & 50$^{\circ}$S-50$^{\circ}$N & 18$^{\circ}$S & 470 & 1000-1200 & 6.65 & 0.0222 & 2.391  \\
\textbf{$\sharp$Glob4*} & 800 & 200 & 0.7 & 20$^{\circ}$S-20$^{\circ}$N & 10$^{\circ}$S 10$^{\circ}$N, less ice at the equator & -5200 & 250-400 & 6.10 & 0.0258 & 2.190  \\
\textbf{$\sharp$Glob5} & 800 & 500 & 0.7 & 30$^{\circ}$S-30$^{\circ}$N & Equator & -2040 & 650-750 & 25.05 & 0.0320 & 1.932  \\
\textbf{$\sharp$Glob6} & 800 & 1000 & 0.7 & 50$^{\circ}$S-50$^{\circ}$N & Equator & 450 & 1200 & 12.00 & 0.0533 & 1.532  \\
\textbf{$\sharp$Glob7} & 1200 & 200 & 0.7 & 20$^{\circ}$S-20$^{\circ}$N & 10$^{\circ}$S 10$^{\circ}$N, less ice at the equator & -5010 & 200-350 & 7.40 & 0.0372 & 1.373  \\
\textbf{$\sharp$Glob8*} & 1200 & 500 & 0.7 & 30$^{\circ}$S-30$^{\circ}$N & Equator & -2250 & 700-800 & 26.10 & 0.0461 & 1.373  \\
\textbf{$\sharp$Glob9} & 1200 & 1000 & 0.7 & 50$^{\circ}$S-50$^{\circ}$N & Equator & 450 & 1200 & 12.30 & 0.0669 & 1.170  \\
\textbf{$\sharp$Glob10} & 400 & 500 & 0.4 & 30$^{\circ}$S-37.5$^{\circ}$N & 5$^{\circ}$N or locally in the depressions & -1740 & 600-800 & 8.70 & 0.8429 & 88.487 \\
\textbf{$\sharp$Glob11} & 800 & 500 & 0.4 & 30$^{\circ}$S-37.5$^{\circ}$N & Equator or locally in the depressions & -1830 & 600-800 & 7.90 & 1.8219 & 70.011 \\
\textbf{$\sharp$Glob12*} & 1200 & 500 & 0.4 & 30$^{\circ}$S-30$^{\circ}$N & Equator or locally in the depressions & -1860 & 700-800 & 7.95 & 2.5917 & 62.722 \\
\hline
\end{tabular}
\end{tiny}
\end{sidewaystable}

\section{Discussion}
\label{secpaleo:discussion}

In this paper we do not seek to reproduce precisely how the SP basin filled with N$_2$ ice, since many parameters are unknown (e.g. the origin of the basin, or the obliquity and the orbital conditions at the time it formed). In addition we do not take into account the reorientation of the rotation axis \citep{Kean:16} and, last but not least, the methane and CO cycles and the presence of methane and CO ices which can strongly affect the surface albedo, emissivity, temperatures and the rheology of the N$_2$ ice (and its sublimation if CH$_4$-rich ice layers form on the N$_2$ ice).
Instead, we seek to evaluate if N$_2$ deposits outside SP could have remained for a long time on Pluto and form perennial deposits, and if yes, at which latitudes.  

In our previous paper \citet{BertForg:16}, we showed that any condensed N$_2$ ice on Pluto's surface tends to end inside the Sputnik Planitia basin. Here, we reproduced similar simulations by taking into account large reservoirs of N$_2$ ice able to sublimate, condense and flow over several Myrs through the changes of obliquity and orbital parameters of Pluto. We found again that any large N$_2$ ice deposits outside SP would accumulate in SP and fill the basin with several kilometres of ice. However, this would take several tens of Myrs during which transient states exist for the deposits. 
Indeed, assuming that the basin formed initially without N$_2$ ice inside, our results show that large deposits of several hundreds of meters of N$_2$ ice, placed at the poles, are not stable there, and would inevitably accumulate first at mid-latitudes over an entire latitudinal band after few Myrs, and then, in some cases, in more equatorial regions after tens of Myrs. We estimate that the basin would be filled up by several kilometers of ice in few Myrs. In the mid-latitude and equatorial regions, the deposits are relatively stable and may remain there during 10-100s of Myrs before to end in Sputnik Planitia, depending on the thermal inertia, the albedo of the ice, the local topography, etc. These results raise discussions about the impact of such glaciers outside SP on the geology of Pluto and on the surface pressures encountered in Pluto's past. 

First, parts of the equatorial regions of Pluto (and in particular in Cthulhu region) are covered by numerous geologically old craters which do not seem particularly eroded by ancient deposition of N$_2$ ice. 
Is it possible that nitrogen ice accumulated in this region and did not eroded the bedrock? We believe that cold/dry based glaciation has a good erosive ability on Pluto and therefore that it unlikely that hundreds of meters of ice accumulated in this region in the past.
Although the erosive properties of nitrogen ice at these temperatures are unknown (nothing has been published on this issue yet), we are guided (1) by the erosive mechanisms that exist on the Earth, where dry/cold based glaciation has been shown to be possible \citep{Atki:02}, although it is difficult to show if it is efficient or not under Pluto's conditions, and (2) by the fact that the water ice bedrock has been strongly eroded around Sputnik Planitia, possibly involving dry glaciation (since on the edges of the ice cap, the ice layer is thin and dry basal flow should dominate).

In addition, if large N$_2$ deposits existed outside SP, they may not have been large enough to flow toward the equatorial regions (like in our simulations started with a global reservoir of ice less than 200~m). Or, the equatorial regions may have been already warmer than the higher latitudes, due to a low thermal inertia (less than 800~SI) or due to albedo gradients (dark tholins at the equator and bright methane ice at higher latitudes, see also \citep{Earl:18}).
    

In some of our simulations, relatively stable deposits of N$_2$ ice are obtained outside SP at higher latitudes, without any ice at the equator. This is for example the case for simulations with a reservoir of 200~m, such as $\sharp$Polar1, $\sharp$Polar4, where the ice does not flow toward the equator but forms latitudinal bands between 25$^{\circ}$S-45$^{\circ}$S and 25$^{\circ}$N-45$^{\circ}$N. 
Interestingly, several surface features on Pluto have been interpreted as evidence for past liquid flow, and they are all observed around the latitudes $\pm$ 30-60$^{\circ}$ \citep{Ster:17}. These latitudes correspond to the regions where ice accumulates in our model (outside SP), first as a transient state (thick glaciers over latitudinal bands) and then as a final state at mid-latitudes around 10-30$^\circ$N (swallower glaciers because most of the ice is trapped inside SP), in particular if a low TI and N$_2$ reservoir are considered. It has been suggested that epochs with higher atmospheric pressure occurred in Pluto's geologic past and enabled the nitrogen ice to be much warmer, perhaps even to turn to liquid, and to flow on the surface leading to the formation of these features \citep{Ster:17}. However, here our simulations demonstrate that surface pressures higher than 100~Pa are unlikely to have occurred in Pluto's past, because the large reservoirs of nitrogen ice are located in the cold and stable equatorial regions and because the relatively high thermal inertia and albedo of the ice limit the sublimation and condensation fluxes. We even found that pressures are the lowest during the high obliquity periods, because N$_2$ ice is never stable at the poles at the scale of the astronomical cycles and therefore not available for intense polar sublimation during these periods. Note that in 2015, which corresponds to northern spring on Pluto, the observed surface of the north polar regions of Pluto was free of N$_2$ ice \citep{Schm:17,Grun:16}. There will be no N$_2$ ice available at the pole to sublimate during summer and increase the pressure. In addition, in our simulations, the maximum surface pressures raise up to $\sim$100 Pa if the ice albedo is set to 0.4, which is a very low value for an ice as mobile as N$_2$ on Pluto. 
Therefore we propose that the paleoliquids - and other terrains thought to have been shaped and altered by liquid flows - are the results of past liquid flows which occurred at the base of massive nitrogen glaciers (basal flow), which accumulated at the mid-latitudes because they are the coldest points on Pluto in average (an effect depending on thermal inertia, as shown by \autoref{tempmean}). These glaciers may have remained at these latitudes for millions of years before they disappear, the ice ending inside SP. 

What would trigger the formation of perennial N$_2$ ice deposits on Pluto outside SP ?
Since the astronomical cycles of Pluto are relatively stable, we can make the hypothesis that the perennial ice deposits on present-day Pluto reached a steady-state. In that case the entire reservoir of N$_2$ ice should be trapped in SP, as suggested by our model results showing that N$_2$ ice deposits outside SP still accumulates in SP after 30 Myrs, losing about 10~m per Myrs. However, other processes could help to maintain perennial N$_2$ ice deposits and feed Pluto's surface with N$_2$ ice outside SP, such as cryovolcanism or bright methane deposits enabling N$_2$ ice to condense on it (see below). 

What is the nature (seasonal or perennial) of the different reservoirs of N$_2$ ice observed in 2015 by New Horizons? How did they form? 
Observationally, it is difficult to know because we do not know the thickness of these reservoirs, although they do not look like several hundreds of meters deep. 
The amounts of diluted CH$_4$ and CO vary in these deposits, which is indicative of volatile evolution processes \citep{Prot:17,Schm:17}.
In \citet{BertForg:16}, we show that regions covered by dark tholins do not favour N$_2$ condensation on it, while surfaces covered by bright methane frost do. In fact, the latitudinal band of nitrogen observed by New Horizons between 30$^{\circ}$N-60$^{\circ}$N has been reproduced by the volatile transport model when high methane albedo ($>$~0.65) were considered (see Figure 3 in \citet{BertForg:16}). In this scenario, the latitudinal band of N$_2$ ice is seasonal since it forms on the cold methane polar frost in winter and sublimates during spring from the pole. 
However, if the thermal inertia is lower than the 800~SI assumed in this scenario, then our results suggest that the ice may be more stable at these latitudes and the latitudinal band of N$_2$ ice may be perennial, continuously fed by seasonal frosts. In other words, bright methane frosts may have helped to maintain the latitudinal bands of massive N$_2$ deposits as a perennial reservoir (e.g. as the one obtained in the case $\sharp$Polar1 or $\sharp$Polar4 in \autoref{uniform1}).
Similar arguments apply for the region East of Tombaugh region: bright methane deposits coupled with relatively low-altitude terrains may favour the accumulation of N$_2$ ice there, which can remain relatively stable over time, especially if the TI is high and thus favouring more stable deposits close to the equator. 
In this paper, thin seasonal polar nitrogen frosts have been obtained in most of the simulations. Although we noted that a lower thermal inertia favour thicker deposits at the poles, simulations taking into account bright methane deposits are necessary to fully investigate the evolution of polar frosts, and will be the topic of future studies.  

Finally, as predicted by the model, N$_2$ ice is more stable in the depressions than in higher terrains. In fact, a limited number of spots of N$_2$-rich ice have been observed in the dark equatorial region of Cthulhu, in particular in the Oort and Edgeworth craters \citep{Schm:17}. 
Note that preferential deposition of N$_2$ ice at the latitudes $\pm$10$^{\circ}$ or $\pm$20$^{\circ}$ (with the equator free of ice) would be consistent with our results obtained with TI between 400-800 SI showing latitudinal bands of stable deposits at these latitudes ($\sharp$Polar1, $\sharp$Equa4, $\sharp$Equa7, $\sharp$Glob4, $\sharp$Glob7). The lack of data makes it difficult to assess, and low-resolution data in the sub-Charon hemisphere is currently under processing and analysis. 
Ground-based telescopic observations rule out the presence of large expanses of N$_2$ ice on the sub-Charon hemisphere, but not the presence of small patches, which are impossible to see from the ground \citep{Grun:13}.

\section{Conclusions}
\label{secpaleo:conclusions}

The Pluto volatile transport model has been used to investigate the cycles of nitrogen on Pluto over diurnal, seasonal and astronomical timescales, taking into account the changes of obliquity, longitude of perihelion and eccentricity and the flow of N$_2$ ice and the changes of topography induced (following the rheology and glacial flow equations as described in \citet{Umur:17}).

Our first conclusion is that Pluto's climate is impacted by the universal Milankovitch mechanism, as the Earth, Mars and Titan. The changes of obliquity and orbital parameters lead to differences of surface temperatures between poles and equator, and asymmetries in the season. We described in this paper how the most volatile ice of Pluto, N$_2$ ice, is impacted by these changes over time. 

We first focused on the nitrogen cycles within the Sputnik Planitia basin, considering that it is the only known perennial reservoir of nitrogen ice on Pluto. 
The results suggest that Sputnik Planitia has a complex history, related to sublimation, condensation, and glacial flow involved at different timescales. 
High obliquity periods induce intense polar summers and thus intense sublimation rates in the northern part of the ice sheet. During the last 2 million years, this part would have lost up to 1 km of ice by sublimation. On the other hand, low obliquity periods favour sublimation in the center of Sputnik Planitia and condensation at the north and south extremities, of up to 300 m of ice in 1 million years.
The glacial flow activity (ice flowing toward the center of Sputnik Planitia) observed at the eastern edge of the ice sheet can thus be related to the intense condensation of nitrogen ice which occurred at these latitudes during the past 2 million years, while the methane-enriched N$_2$ ice dark plains are linked to the intense sublimation which occurred north of Sputnik Planitia during the same period. The deep pits observed in the south of Sputnik Planitia may have started to form 100,000 years ago, when the southern latitudes of the ice sheet entered a net sublimation-dominated regime. The bright plains in the center of Sputnik Planitia can be explained by the current seasonal accumulation of ice there. Finally, the depressions observed north and south of the ice sheet, as well as the strong erosion of the Al-Idrisi Montes, are consistent with the simulated glacial activity of Sputnik Planitia, with continuous variation of elevations at the edges of the ice sheet up to 300 m every obliquity cycle. The results also show that in current epoch, the ice sheet is close to its minimal extension (in the model in current epoch the center of SP has a higher elevation than the northern and southern edges of SP), which is consistent with the observations showing evidences of strong erosion further north (Al-Idrisi) and south (West and East of Tenzing Montes) of the ice sheet.

We also explored the stability of N$_2$ ice deposits outside Sputnik Planitia. Our simulations show that nitrogen ice tends to end inside Sputnik Planitia but if large deposits are formed outside SP, they should accumulate and persist in the mid-latitude and equatorial regions for several tens of million years. In particular, N$_2$ ice accumulates in the depressions. For instance, in most of the simulations involving N$_2$ ice in the equatorial regions, no N$_2$ ice has been obtained in the Tartarus Dorsa region, featuring the high altitude bladed terrains.  
The latitudes where N$_2$ ice accumulates depends on the seasonal thermal inertia (the higher it is, the more equatorial are the deposits), the ice albedo, the initial distribution and probably other parameters not taken into account in this paper such as the methane ice distribution.   
Our simulations support the case of low to medium thermal inertia (400-800 SI) for several reasons. It enables to reproduce the evolution of pressure since 1988 \citep{BertForg:16}. In some cases, it enables formation of perennial deposits at mid-latitudes but not at the equator, which remains free of volatile ice. 

Geomorphological evidences of past liquid flows have been observed at Pluto's surface at the same mid-latitude. Therefore we suggest that they formed by liquid nitrogen flows at the base of ancient thick nitrogen glaciers instead of formed by liquid nitrogen flows directly at Pluto's surface during higher pressure epochs in Pluto’s geologic past, as suggested by \citet{Ster:17}.
This is reinforced by our results showing that the minimum and maximum surface pressures obtained in our simulations always remain in the range of milli-Pascals and tens of Pascals, respectively. Therefore surface temperatures never reach the triple point of nitrogen. It is not possible to reach higher pressures in Pluto's past with our model because the sublimation-condensation flux are limited by the medium to high thermal inertia and the relatively bright albedo assumed for the N$_2$ ice ($>$ 0.4). 

Finally, the cycle of nitrogen ice on Pluto can be impacted by other processes, not taken into account in the simulations of this paper. In particular, methane ice is known to play a complex role since it can cold trap nitrogen ice if its albedo is high enough \citep{BertForg:16,Earl:18}, which could explain why many patches of nitrogen ice are observed outside Sputnik Planitia. A study taking into account both cycles of methane and nitrogen, over all timescales, is in preparation and should help to better understand how these ices evolve on Pluto.

\newpage

\bibliographystyle{plainnat}
\bibliography{biblio}


\ack
We acknowledge the Centre National d'Etudes Spatiales (CNES) for its financial support through its ``Syst\`eme Solaire`` program. The authors thank the whole NASA \textit{New Horizons} instrument and scientific team for their excellent work on a fantastic mission and their interest in this research.
\label{lastpage}


\clearpage	

\end{document}